\documentclass[letterpaper,11pt]{article}
\usepackage{amsmath, epsfig, amssymb,array}

\begin{document}

\newcommand{\bref}[1]{(\ref{#1})}
\def\sl#1{\mathord{\not\mathrel{{\mathrel{#1}}}}}
\def\ol#1{\overline{#1}}
\def\eps{\epsilon}
\def\rvh{r_{vh}}
\def\dy{{\delta y}}
\def\ymi{{y_{\rm min}}}
\def\beq{\begin{eqnarray}}
\def\eeq{\end{eqnarray}}
\def\be{\begin{equation}}
\def\ee{\end{equation}}
\baselineskip=18pt

\newcommand{\bear}{\begin{array}}
\newcommand{\ear}{\end{array}}

\newcommand{\beqa}{\begin{eqnarray}}
\newcommand{\eeqa}{\end{eqnarray}}
\newcommand{\no}{\nonumber}
\def\OMIT#1{{}}
\newcommand{\lsim}{\mathrel{\rlap{\lower4pt\hbox{\hskip1pt$\sim$}}
    \raise1pt\hbox{$<$}}}         
\newcommand{\gsim}{\mathrel{\rlap{\lower4pt\hbox{\hskip1pt$\sim$}}
    \raise1pt\hbox{$>$}}}         
\newcommand{\order}{\cal O}


\thispagestyle{empty}
\vspace{20pt}
\font\cmss=cmss10 \font\cmsss=cmss10 at 7pt

\begin{flushright}
\today \\
UMD-PP-09-063 \\
\end{flushright}

\hfill
\vspace{15pt}

\begin{center}
{\Large \textbf { Astrophysical Implications  of a Visible Dark Matter Sector from
a  Custodially Warped-GUT
} }
\end{center}

\vspace{11pt}

\begin{center}
{\large Kaustubh Agashe$\, ^{a}$,
Kfir Blum $\, ^{b}$, Seung J. Lee$\, ^{b}$ and Gilad Perez$\, ^{b}$} \\
\vspace{15pt}
$^{a}$\textit{Maryland Center for Fundamental Physics,
     Department of Physics,
     University of Maryland,
     College Park, MD 20742, USA}
\\
$^{b}$\textit{Department of Particle Physics \& Astrophysics, Weizmann Institute of Science,
Rehovot 76100, Israel}
\end{center}

\vspace{14pt}

\begin{center}
\textbf{Abstract}
\end{center}
\vspace{4pt} {\small \noindent

We explore, within the warped extra dimensional framework, the possibility of finding anti-matter signals in
cosmic rays (CRs) from dark matter (DM) annihilation. We find that exchange of order 100 GeV radion, an
integral part of this class of models, generically results in a sizable Sommerfeld enhancement of the
annihilation rate for DM mass at the TeV scale. No ad-hoc dark sector is required to obtain boosted
annihilation cross sections and hence signals. 
Such a 
mild hierarchy between the radion and DM masses can be natural
due to the pseudo-Goldstone boson nature of the radion. We study the implications of a Sommerfeld enhancement
specifically in warped grand unified theory (GUT) models, where proton stability implies a DM candidate. We
show, via partially unified Pati-Salam group, how to incorporate a custodial symmetry for $Z\to b\bar b$ into
the GUT framework such that a few TeV Kaluza-Klein (KK) mass scale is allowed by electroweak precision tests.
Among such models, the one with the smallest $SO(10)$ (fully unified) representation, with $SU(5)$ hypercharge
normalization, allows us to decouple the DM from the electroweak gauge bosons. Thus, a correct DM relic density
can be obtained and direct detection bounds are satisfied. Looking at robust CR observables, we find a possible
future signal in the $\bar p / p$ flux ratio consistent with current constraints. Using a different choice of
representations, we show how to embed in this GUT model a similar custodial symmetry for the right handed tau,
allowing it to be strongly coupled to KK particles. Such a scenario might lead to observed signal in CR
positrons; however, the DM candidate in this case can not constitute all of the DM in the universe. As an aside
and independent of GUT or DM model, the strong coupling between KK particles and tau's can lead to striking LHC
signals. }

\vfill\eject \noindent

\tableofcontents


\section{Introduction}

In the last few years a host of experiments have provided us with detailed cosmic ray (CR) data in the energy
range of 10-1000
GeV~\cite{Adriani:2008zr,:2008zzr,Abdo:2009zk,Collaboration:2008aaa,DuVernois:2001bb,Boezio:2008mp,Alcaraz:2000bf,Barwick:1997ig,Torii:2001aw,Torii:2008xu,
Aharonian:2009ah}. The data is interesting for the astrophysics and cosmology communities, enabling them to
learn about production and propagation of particles in the Galaxy. It is also of great interest for the
particle physics community, due to the anticipation that annihilation of dark matter (DM), possibly consisting
of weakly interacting massive particles (WIMPs), would generate an observable signal in the CR data.
A lot of model building effort has recently been associated with the PAMELA~\cite{Adriani:2008zr} and
ATIC/FERMI/HESS~\cite{:2008zzr,Abdo:2009zk,Collaboration:2008aaa,Aharonian:2009ah} measurements. Probably the
main reason for the excitement is due to a rise in the positron to the total electron flux ratio (positron
fraction) in the 10-100\,GeV energy range, as measured by PAMELA. The rising positron fraction is in tension
with common assumptions regarding the production and propagation of CR electrons and positrons in the Galaxy
(see {\it e.g.}~\cite{Morselli:2008uk, Delahaye:2008ua}, and references within).

The rising positron fraction\footnote{
See, however, \cite{Schubnell:2009gk} for cautionary notes.}, though certainly intriguing,
does not necessarily imply an ``anomaly" with respect to what could be expected from standard astrophysics as follows: The
actual positron intensity does not exhibit an excess when contrasted with model independent
calculations~\cite{Katz:2009yd}, which successfully describe the observed abundance of other secondary CR particles, such as antiprotons. 
Moreover, since measurements of unstable CR isotopes can be used to infer the
cooling suppression of positrons at an energy of around 20\,GeV, a theoretical estimate for the corresponding positron flux can be derived at that energy~\cite{Katz:2009yd}. Thus, the combination of
the predicted positron flux and the available $e^-+e^+$ data~\cite{:2008zzr,Abdo:2009zk,DuVernois:2001bb,Alcaraz:2000bf,Barwick:1997ig,Torii:2001aw} yields an independent  
estimate of the background positron fraction for this energy range.
The authors of~\cite{Katz:2009yd} have compiled the above data and shown that it is, in fact, consistent with the PAMELA measurement, leaving little room for an anomaly.
It is, therefore, conceivable that the rising positron fraction may
just imply that the currently fashionable diffusion models for CR propagation in the Galaxy are incorrect.

Even within simple diffusion models, the PAMELA result has been argued to be compatible with secondary positrons, provided that the primary electron spectrum is
soft~\cite{Delahaye:2008ua,Delahaye:2007fr}.
%
Along these lines there are alternative astrophysical interpretations, wherein the positrons are still of secondary
origin~\cite{Blasi:2009hv,Cowsik:2009ga,Shaviv:2009bu}. We note that, at present, all of these
astrophysical interpretations require further assessment in order to verify the compatibility of the rising
positron fraction with the CR nuclei and antiproton data. In regards to suggested primary injection mechanisms,
pulsars have been put forth as astrophysical source candidates (see {\it e.g.}~\cite{pulsars}), and models of DM
annihilation or decay have been proposed as a particle physics explanation (see {\it
e.g.}~\cite{Cholis:2008wq,Cirelli:2008pk,ArkaniHamed:2008qn,Hooper:2009zm,He:2009ra}).

A common feature of the DM annihilation models which can account for observable contributions of anti-matter
CRs is the presence of a large enhancement (``boost'') factor in the annihilation cross section.
This feature can
be traced back to the WMAP data, which fixes the annihilation cross section at the cosmological epoch to
$\langle\sigma v\rangle \sim\mathrm{few} \ 10^{-26}{\rm \,cm^3\,s}^{-1}$.
For DM mass in the TeV range, the latter number implies that the positron (or other anti-matter particle) injection rate lies orders of magnitude bellow the astrophysical background.
One widely studied possibility
for obtaining a boost factor of the velocity weighted current annihilation
cross section relative to its cosmological value at freeze-out
is the so-called ``Sommerfeld Enhancement'' (SE)~\cite{Somm},
which originates from DM particles interacting via a light force carrier.
Other forms of enhancement have also been studied in the literature~(see~{\it e.g.}~\cite{Feldman:2008xs, Ibe:2008ye,Hooper:2008kv}).

While not currently necessitated by data, it is still an interesting possibility that the observed CR fluxes include a primary component from DM annihilation. Furthermore, in wait of future data release by the PAMELA and upcoming missions~\cite{Alpat:2003pn,Ahn:2008my}, it is timely to consider theoretically clean observables for indirect detection, such as the antiproton to proton, and positron to antiproton flux ratios~\cite{Katz:2009yd}.

In this paper, we explore such robust observables in the future
measurements using a well motivated theoretical framework,
namely, that of a warped extra dimension a la Randall-Sundrum model (RS1)~\cite{rs1}, but with SM fields propagating in it.
One nice feature of the warped extra dimension framework in light of indirect astrophysics signal is that there is a natural candidate for the force carrier of SE, namely the radion, which is an intrinsic component of the
theory.\footnote{Based on AdS/CFT correspondence
this nice feature of radion as a mediator of SE
is {\em dual} to dilaton exchange
in 4D CFT theories of
electroweak symmetry breaking with appropriate DM candidate.
Reference~\cite{Bai:2009ms} considered dilaton as messenger between
SM and dark sector, but did not study the SE from dilaton exchange.
}
It is the degree of freedom corresponding to fluctuations of the size of the extra dimension in an RS-type scenario.
Radion mass is in principle
a free parameter of the theory, but assuming no fine tuning
(and Kaluza-Klein (KK) scale of ${\cal O}(3\rm\,TeV)$)
its mass could
vary from ${\cal O}(100\rm\,GeV)$ -- in which limit
it can be consider as a pseudo Goldstone boson (PGB)\footnote{However, unlike other PGB's, the radion can have sizable non-derivative couplings (required
for Sommerfeld enhancement) even in the GB limit.}-- all the way to the KK
scale.
The precise radion mass depends on
%
%
mechanism which stabilizes the distance between the
branes~\cite{Goldberger:1999uk, Goldberger:1999un, CGRT, DeWolfe:1999cp, TanakaMontes, Csaki:2000zn,Garriga:2002vf}.
Also, the
radion coupling to other particles are (roughly) given by mass
of the other particles in units of the KK
scale.
Hence, for a TeV scale DM, radion coupling to DM pair is ${\order}
\left( 1 \right)$, and radion mass as large a
few hundred GeV can give a significant SE.

We focus here on a variant of the DM model based on a
grand unified theory (GUT)
model within this framework~\cite{Agashe:2004ci}\footnote{Based on above
discussion, it is clear
that radion mediated SE might also be relevant for
{\em other}
RS-type scenarios with DM~\cite{Agashe:2007jb,otherRSDM}.},
where stability of the DM is a spin-off of
suppressing proton decay.
The
DM particle in this model is a SM gauge singlet
and is a non-standard GUT partner of the top quark.
We incorporate custodial symmetry protection of $Z$ coupling to
bottom quarks~\cite{custodial2} into the above existing
RS-GUT model in order to suppress
otherwise large shift in this coupling, and construct
several models of this type. For simplicity, we mainly focus on
partial
unification
based on the Pati--Salam group, which captures the major experimental implication; however, full unification is discussed as well.

We also explore the consequences of implementing
a similar custodial symmetry protection of $Z$ couplings to right-handed
(RH) tau's in order to accommodate
the possibility of
RH tau's
%
%
being localized near the TeV end of the extra dimension
and hence
having a large coupling to KK particles
in this model.
In such a scenario,
DM annihilation can have a large leptonic branching ratios (BRs)
via $Z^{ \prime }$ -- the extra $U(1)$ of Pati-Salam -- exchange.\footnote{Recall that
the DM is a SM singlet so that KK exchange of
SM gauge fields is not allowed at leading order.}
It is interesting that
such large leptonic BRs can result in indirect detection in
CR positron/electrons.
%
%
We emphasize that, independent of GUT or DM model,
such a possibility,
in turn, opens up new doors for searching for
KK particles (for example,
KK $Z$) at the LHC through their decays highly-boosted RH taus, which will be
a relatively clean signal with negligible Standard Model (SM) background.

We find, however, that models with significant DM-$Z'$ couplings which
allow for such an exciting astrophysics phenomenology, in general,
yield a too small primordial DM density or
are
possibly in tension with direct detection bounds.
Furthermore,
this scenario seems to
require very large representations when
fully unified into $SO(10)$
and in any case, it
is incompatible with
$SU(5)$ normalization of hypercharge.
Thus, even the SM-level of unification
of gauge couplings (which is automatic in warped models with
$SU(5)$ normalization of hypercharge~\cite{SMlevelunif}) is not
guaranteed to be
maintained.

We hence consider other class of models, which can be fully unified into
not-so-large $SO(10)$ representations, and furthermore preserves the
$SU(5)$ normalization of hypercharge.
%
%
Thus,
SM level of unification of gauge couplings is maintained and 
even unification with a precision comparable to the supersymmetric SM one might be possible as
in reference~\cite{Agashe:2005vg}.
It is quite interesting
that this model
actually predicts vanishing DM-$Z'$ coupling so that the above constraints
from
relic density and direct detection
are
all satisfied, albeit (as a corollary)
not leading to exciting astrophysics signals
in positron/electron channel.
Note, however, that custodial $Z\to \tau\bar \tau$ symmetry protection
can still be implemented in this GUT model so that that the exciting
LHC phenomenology associated with tau's is possible.

The outline of the rest of the paper
is as follows. We begin in section \ref{model} with a description of the model (with more details
in appendix) which is a modified version of the warped extra dimensional DM model of references
\cite{Agashe:2004ci}. In section \ref{annihilation}, we discuss implications for cosmology and astrophysics.
We explore the SE arising in our framework with a (light) radion. We proceed
to calculate the DM relic density and direct detection cross-sections.
The parameter space
compatible with WMAP observations and CDMS bounds is delimited.
A set of benchmark models within this allowed parameter space is defined,
in which a large SE factor is a natural consequence of the setup.
Then, we discuss both particle and astro-physics aspects of DM annihilation.
Results of our detailed analysis are presented.
In section~\ref{sec:coll} we briefly discuss the radion-related collider signals at the LHC
with conclusions are drawn in section \ref{conclude}.

\section{The Model}
\label{model}

We first present a review of the general warped extra dimensional
framework and then of the DM model within it.
For a review and further references, see the reference~\cite{Davoudiasl:2009cd}.
The reader interested only in the particle content of the model and the couplings
relevant for signals in cosmic ray experiments
can skip to tables \ref{content}, \ref{couplingscust1} and \ref{couplingscustGUT}
and the comments listed there.

\subsection{SM fields in bulk of warped extra dimension}

The Randall-Sundrum (RS1) framework consists of a slice of anti-de Sitter
space in five dimensions (AdS$_5$), where
the warped geometry
naturally generates the Planck-weak hierarchy as follows~\cite{rs1}.
The $4D$
graviton, i.e., the zero-mode
of the $5D$ graviton, is automatically localized
at one end of the extra dimension (hence called the Planck/UV brane).
If the Higgs sector is localized at the other end (hence
called the TeV/IR brane)
\footnote{In fact
with SM Higgs originating as 5th
component of a $5D$ gauge field
($A_5$) it is automatically so~\cite{Contino:2003ve}.},
then the
UV cut-off for quantum corrections to the Higgs
mass can be $\sim \left( \hbox{TeV}
\right)$, whereas simultaneously the $4D$ gravitational coupling strength being
set by the usual Planck scale,
$M_{ \rm Pl} \sim 10^{ 18 }$ GeV. Such a hierarchy of
mass scales at the two ends of the extra dimension is stable against
quantum corrections in the
warped geometry, where the effective $4D$ mass scale (including
UV cut-off) is dependent
on position in the extra dimension.
Specifically, TeV $\sim M_{\rm Pl}\,
e^{ - k \pi R }$, where
$k$ is the AdS$_5$ curvature scale and $R$ is the proper
size of the extra dimension. The crucial
point is that the required
modest size of the radius (in units of the curvature radius), i.e.,
$k R \sim 1 / \pi \log \left( M_{ \rm Pl}  / \hbox{TeV}
\right) \sim 10$ can be
%
%
stabilized with only a corresponding
modest tuning in the fundamental or $5D$ parameters of
the theory~\cite{Goldberger:1999uk, Garriga:2002vf}.
Remarkably, the correspondence between
AdS$_5$
and $4D$ conformal field theories (CFT)~\cite{Maldacena:1997re}
suggests that
the scenario with warped extra dimension is
dual to the idea of a composite Higgs in $4D$
\cite{Arkani-Hamed:2000ds, Contino:2003ve}.

In the original RS1 model, it was assumed that the rest of the SM, i.e.
gauge and fermion, fields are also localized on the TeV brane (just like
the Higgs). Such a scenario does not have a built-in explanation for
the hierarchy between quark
and lepton masses and mixing angles (flavor
hierarchy). In addition, the scenario
generically also has a flavor and
proton stability problems as follows. The (effective) cut-off for the
{\em entire} SM (i.e., not just the Higgs) is of ${\cal O}
\left( \hbox{TeV} \right)$
in this case so that the higher-than-dimension-$4$
SM operators induced
by the UV completion of RS1 will lead to too large
flavor changing neutral currents (FCNC's) and too rapid proton decay:
recall that such operators have to be suppressed by, at least,
${\cal O}\left(10^5\right)$ TeV (if they violate CP in addition)
and $\sim 10^{ 15 }$ GeV, respectively, to be consistent with the data.
The above argument suggests that similar problem would be present for the electroweak (EW)
sector, a manifestation of the little hierarchy problem.

\subsubsection{Solution to flavor puzzle and problem}

It was realized that with
SM fermions propagating in the bulk, i.e.,
arising as zero-modes of $5D$ fermions,
we can
account for the flavor hierarchy as well
\cite{gn, gp}.
The idea is
that the effective $4D$ Yukawa couplings of the SM fermions are given
by a product
of the fundamental $5D$ Yukawa couplings and
the overlap of the profiles (of the SM fermions and
the Higgs) in the extra dimension.
Moreover, vastly different
profiles in the extra dimension for the SM fermions
and hence their hierarchical
overlaps with Higgs,
can be easily obtained by small variations
in the $5D$ fermion mass parameters.
Thus, hierarchies in the $4D$ Yukawa couplings
can be generated without any (large) hierarchies in
the fundamental $5D$ parameters ($5D$ Yukawa couplings
and $5D$ mass parameters for fermions).

As a bonus, the above-mentioned flavor problem is also
solved as follows.
Based on the above discussion, we can see that
light SM fermions are chosen to be localized near Planck brane in such
a way that the
effective cut-off for them is $\gg$ TeV.
In more detail (this discussion will be useful in what follows),
the contribution of cut-off effects is actually
dominated by near the TeV brane (where
the effective cut-off is of course of ${\cal O}\left(\hbox{TeV}\right)$), but
the operators are further suppressed by profile
of the SM fermions near the TeV brane.
Since the same profiles dictate the $4D$ Yukawa coupling,
we see that $4$-fermion operators have
a coefficient $\sim1 / \hbox{TeV}^2 \times (4D \; \hbox{Yukawa} )^2$
which is
sufficient to suppress FCNC's:
\begin{eqnarray}
\begin{array}{c}
\psi_{ SM }^4 / \hbox{TeV}^2  \\
(\hbox{SM on TeV brane})
\end{array}
& \rightarrow & \begin{array}{c}
\psi_{ SM }^4 / \hbox{TeV}^2 \times \hbox{profiles at TeV brane}
\\
(\hbox{SM in bulk})
\end{array} \nonumber \\
& \sim & \begin{array}{c}
\psi_{ SM }^4 / \hbox{TeV}^2 \times ( 4D \; \hbox{Yukawa} )^2
\end{array}
\end{eqnarray}

As a corollary, the SM gauge fields must also propagate in the bulk
(hence the scenario is called ``SM in the bulk''). Thus,
the couplings of SM fermions (with
different profiles) to gauge KK
modes
are non-universal, resulting in flavor violation
from exchange of these KK modes~\cite{Delgado:1999sv}.
However, there is a built-in analog of GIM mechanism of the SM in
this framework~\cite{gp,hs, aps} which suppresses FCNC's.
Namely, the non-universalities
in couplings of SM fermions to KK modes are of the size of $4D$ Yukawa
couplings since KK modes have a similar profile to the SM Higgs, i.e.,
gauge KK modes are localized near the TeV brane. Thus,
even though the gauge KK mass is of $ {\cal O}
\left( \hbox{TeV} \right)$, FCNC's from their exchange
can be adequately suppressed.\footnote{A residual ``little CP problem"~\cite{LCP} is still present~\cite{aps,Problem} in the above scenario which can be amended by various alignment mechanisms~\cite{LCP,Align}.}
Similarly, the KK modes induce effects on electroweak
precision tests (EWPT), which can be brought under control
by suitable imposing custodial symmetries~\cite{custodial1, custodial2}.

\subsubsection{Baryon symmetry}

Satisfying the
constraints from non-observation of
proton decay requires, however, the mass scale of
new physics to be generically of $ {\cal O} \left( 10^{ 15 } \right)\,$GeV
so that Yukawa-type suppression
of cut-off effects on top
of ${\cal O}
\left( \hbox{TeV} \right)$ scale
discussed above
is not enough in this case.
A simple solution is to
impose a gauged\footnote{Global symmetries
are expected to be violated by quantum gravity effects.} baryon-number
symmetry, denoted by
$U(1)_B$, in the bulk and to break it (arbitrarily) on Planck brane
so that the ``would-be'' zero-mode gauge boson is projected out.
Thus,
proton decay operators can originate only on the Planck brane,
where they
are adequately (i.e., Planck-scale which is the cut-off
there) suppressed:
\begin{eqnarray}
\begin{array}{c}
q^3 l/ \hbox{TeV}^2  \\
(\hbox{SM on TeV brane})
\end{array}
& \rightarrow & \begin{array}{c}
q^3 l / M_{Pl.}^2
\\
(\hbox{SM in bulk})
\end{array}
\end{eqnarray}

\subsection{Dark matter from proton stability in GUT}

Extending the bulk gauge symmetry from the SM to a grand unified
theory (GUT) is
motivated by the resulting SUSY-level precision
gauge coupling unification
\cite{Agashe:2005vg}, in addition to an explanation of quantized hypercharges
of the SM fermions.

However, the extra gauge bosons in the GUT -- for example,
$X$, $Y$ in the case of
$SU(5)$ -- have their KK excitations (with a mass of ${\cal O} \left(
\hbox{TeV} \right)$)
localized near the TeV
brane (even if their ``would-be'' zero-modes
can be decoupled by suitable breaking of the GUT).
Hence, if the SM quarks and leptons are {\em grand-unified}
as well, i.e., they arise as
zero-modes from the same $5D$ multiplet in a GUT representation,
then the, $X$ and $Y$ exotic, gauge KK modes will mediate proton decay with only Yukawa
suppression [beyond their ${\cal  O}
\left( \hbox{TeV} \right)$] mass which is clearly not sufficient.

\subsubsection{Split multiplets}

\begin{figure}[hbp]
\begin{center}
\includegraphics[width=150mm]{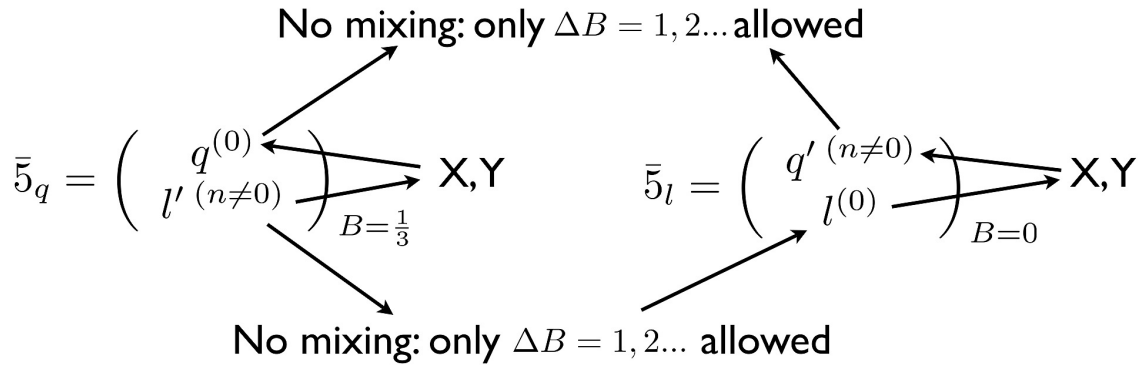}
\caption{Split multiplets}
\label{fig:split}
\end{center}
\end{figure}

The solution is to invoke
``split'' multiplets, namely, we break the GUT
group down to the SM by boundary conditions (on the Planck brane
so that gauge coupling unification still works).
We can then
choose SM quark and lepton to be zero-modes of two {\em different}
$5D$ multiplets in a GUT representation. The extra (i.e., ``would-be'')
zero-modes with SM gauge quantum numbers of
lepton and quark, respectively, from the two $5D$
multiplets can be projected out by the boundary condition, i.e.,
these fields only have only {\em massive} KK excitations.
In this way,
the $X$, $Y$ gauge bosons cannot couple SM quarks
to SM leptons (again such a coupling can only arise if
SM quarks and SM leptons are contained in the same $5D$ multiplet):
see Fig. \ref{fig:split}.

However,
higher-order effects
can ``undo'' the splitting of quark and lepton multiplets
so that proton decay can strike again --
for example, brane-localized mass terms can mix the (KK) leptons from
the ``quark'' multiplet (i.e., which contains a quark zero-mode)
with the zero-mode lepton from the other (lepton) multiplet
and similarly mix (KK) quarks from the lepton multiplet
with the quark zero-mode from the quark multiplet.
In any case,
we still have to contend with cut-off effects giving proton decay.
A simple way out is to impose a $U(1)_B$ gauge symmetry in the bulk
as discussed in the case of non-GUT model. Specifically, the
{\em entire} $5D$ quark (lepton) multiplet,
including the KK leptons (quarks) contained in it,
are assigned $B = 1/3 \; (0)$.


\subsubsection{$Z_3$ symmetry}

Unlike in the non-GUT model,
a ({\em discrete}) subgroup of $U(1)_B$ has to be preserved during
the breaking $U(1)_B$
on the Planck brane in order to prevent mixing between the
KK leptons from quark multiplet with the lepton zero-modes from
the lepton multiplet
(and thus avoid catastrophic
proton decay). For example, it is possible to require that the $U(1)_B$ symmetry
is only broken by scalar fields with integer charges, i.e.,
only $\Delta B = 1, 2,...$
operators are allowed. Thus, the above-mentioned
mixing of ``wrong'' (i.e., KK)
lepton (or quark) with $B=1/3 \; (\rm or\  0)$
with correct zero-mode
with $B = 0 \; (\rm or\ 1/3)$ is forbidden (see Fig. \ref{fig:split}),
even though $4$-fermion
proton decay operators , albeit safe due to the Planckian
suppression, are allowed.

The crucial observation is that,
as a corollary, the GUT partners of the SM quarks and leptons, i.e.,
the (KK) leptons (quarks) from quark (lepton) multiplet
cannot decay into purely SM particles due to their
``exotic'' baryon-number assignment. Explicitly,
the extra particles in the GUT model
(including $X$, $Y$ gauge bosons) are charged under the following
$Z_3$ symmetry
\begin{eqnarray}
\Phi & \rightarrow & e^{ 2 \pi i \left( \frac{ \alpha - \bar{\alpha} }{3}
- B \right) } \Phi
\end{eqnarray}
(where $\alpha$, $\bar{\alpha}$ are the number of color, anti-color
indices on $\Phi$)
whereas the SM particles -- having
correct combination of color and baryon-number are neutral under it.
Thus, the lightest $Z_3$ charged particle (dubbed ``LZP'') is stable.

In references~\cite{Agashe:2004ci}, an $SO(10)$ model with
canonical representations for SM fermions, i.e., in $\bf{16}$ was
presented. It was shown that
SM singlet (RH neutrino) partner of $t_R$\footnote{The $t_R$ multiplet
being the one giving the LZP follows from
its profile being closest to the TeV brane.} can be
the LZP and is in fact a WIMP and therefore a good dark matter candidate:
a spin-off of suppressing proton decay (analogous to $R$-parity in supersymmetry).\footnote{In addition to
DM, other GUT partners could also give interesting signals (see {\it e.g.}~\cite{Agashe:2004ci, Gripaios:2009dq}).}

\section{Partially \& Fully Unified Custodial Models}

In models with the canonical/minimal choice
of EW quantum numbers, the shift in
$Z b \bar{b}$ resulting from
exchange of KK modes is typically (a
bit) larger than that allowed by EWPT.
This shift results in ${\cal O}\rm(5\,TeV)$ lower bound on the KK scale which implies a rather
severe little hierarchy problem.
A custodial symmetry to protect such a shift in $Z b \bar{b}$
was proposed in reference~\cite{custodial2} which
requires non-canonical EW quantum numbers.

Here, we incorporate such a custodial symmetry
in the warped GUT DM model of reference~\cite{Agashe:2004ci},
presenting several models of this type.
For simplicity, we mainly
work with partially unified, i.e., Pati-Salam, gauge group
and
comment on full unification into $SO(10)$ on case by case basis.
%
%
%
%
Note that there is no proton decay from exchange of
$X$, $Y$-type GUT gauge bosons in Pati-Salam model
so that there is no motivation for incorporating split multiplets
and hence for existence of DM of this type
in this case.
However,
we always have full unification into $SO(10)$,
where DM emergence is a spin-off of proton stability
as mentioned above, at the back of our minds.

It is interesting that such a symmetry can also be extended to leptons in order to protect the shift
in $Z$ coupling to leptons. Thus,
leptons (in particular, $\tau$) can be localized closer to the TeV brane,
resulting in larger (than canonical) couplings of
gauge KK modes to $\tau$. A significant DM annihilation to
$\tau$ via exchange of KK gauge bosons, therefore, might be possible,
which may be relevant for the PAMELA rise (or future signals). As further discussed below this possibility is typically in tension with the observed DM relic density and with direct detection limits.

Finally, although we focus here on models where DM is a
SM gauge singlet
GUT partner of $t_R$, it
is worth noting that DM could also be GUT partner of
$(t,b)_L$ instead, depending on details such as
the proximity of these profiles to the TeV brane.
We will defer study of such a possibility to the future.

%
%

\subsection{Canonical}

Just to get oriented, the
canonical choice for representations under Pati-Salam
group, i.e., $SU(4)_C \times SU(2)_L \times SU(2)_R$ are in
table \ref{canonical}.\footnote{Of course, we
can invoke split multiplets so that there can be two
-- one for quarks and one for leptons -- multiplets
of each type in the table.}
Namely, LH SM fermions, i.e., $SU(2)_L$ doublet
quarks and leptons, are $SU(2)_R$ singlets with
$T_{ 3 R } = 0$.
 RH quarks and leptons, i.e., $SU(2)_L$ singlets,
are
$SU(2)_R$ doublets with
$T_{ 3 R } = \pm 1/2$ for RH up quark (or RH neutrino) and RH down
quark (or RH charged lepton).
The SM
hypercharge is then given by
\begin{eqnarray}
Y & = & T_{ 3 R } - \sqrt{2/3} X,
\end{eqnarray}
where $X$ are the charges under the non-QCD $U(1)$ generator
present in $SU(4)_c$, {\it i.e.}, $SU(4)_c \sim SU(3)_c \times
U(1)_X$.
We have chosen $X= \hbox{diag} \sqrt{3/8}
\left( - 1/3, - 1/3, - 1/3, 1 \right)$ when acting on $\bf{4}$
of $SU(4)_c$ such that the normalization
for this generator acting on
$\bf{4}$ of $SU(4)_c$ is Tr$X^2 = 1/2$.
This combination of $T_{ 3 R }$ and $X$
corresponds to the
$SU(5)$ normalization of hypercharge when fully unified into $SO(10)$.
Thus this model (at the least)
maintains the SM-level of
unification of couplings, even in the context of a warped extra dimension.

Thus, we have the breaking
pattern: $SU(4)_c \times SU(2)_R \rightarrow SU(3)_c \times U(1)_Y$
achieved by boundary condition on the Planck brane.
The Pati-Salam group is preserved by boundary conditions on the
TeV brane (of course Higgs VEV breaks
$SU(2)_L \times SU(2)_R \rightarrow SU(2)_V$).
The gauge field that corresponds to the combination of $T_{ 3 R }$ and $X$ which is orthogonal to
hypercharge will be denoted by $Z^{ \prime }$.
The couplings to $Z^{ \prime }$ are then given by
(up to overlap factor denoted below by $a$)
$\left( g_{ R } / \cos \theta^{ \prime }
\right) \left( T_{ 3 R } - Y \sin^2 \theta^{ \prime }
\right)$, where $\sin^2 \theta^{ \prime }
\equiv \left( {3\over 2}g_4^2 \right) / \left( {3\over 2} g_4^2 + g_{ R }^2 \right)$
and $g_{ R}, g_4$ are the ``4D'' couplings of
$SU(4)_C$ and $SU(2)_R$ gauge groups, respectively
(obviously the normalized $U(1)_X$ gauge coupling is
same as the $SU(4)_c$ one).

Note that, due to
Pati-Salam being only partial unification
of SM gauge groups,
the $SU(2)_R$ and $SU(4)_c$ gauge couplings are unrelated
so that $\sin^2 \theta^{ \prime }$ is a free parameter.
However, it was shown in reference~\cite{Agashe:2005vg} that a $SO(10)$-type
completion of Pati-Salam, {\it i.e.}, full unification
of SM gauge groups, is very well-motivated due to the SUSY-level
precision of the
gauge coupling unification. With this result in mind,
we can set $g_4 = g_R$ to find $\sin^2 \theta^{ \prime } = 3/5$.


\begin{table}[hbt]
\begin{center}
\begin{tabular}{|c|c|c|c|}
\hline
& $SU(4)_c \sim SU(3)_C \times U(1)_X$ & $SU(2)_L$ & $SU(2)_L$ \\
\hline
LH & \bf{4} $\sim$ \bf{3}$_{ - \frac{1}{3} } +$
\bf{1}$_{ 1 }$ & \bf{2} & \bf{1}
\\
\hline
RH & \bf{4} $\sim$ \bf{3}$_{ - \frac{1}{3} } +$
\bf{1}$_{ 1 }$ & \bf{1} & \bf{2} \\
\hline \hline
$H$ & \bf{1} & \bf{2} & \bf{2} \\
\hline
\end{tabular}
\end{center}

\caption{Canonical representations for
SM fermions and Higgs: the subscripts denote
the $ \sqrt{8/3} \; X$ charge.}
\label{canonical}
\end{table}


\subsection{Custodial Pati-Salam model}

As outlined above,
we begin by constructing a model with custodial symmetry for
$Z b \bar{b}$ based on partial
unification, namely, the Pati-Salam gauge group:
$SU(4)_C \times SU(2)_L \times SU(2)_R$ we later discuss how to fully unify it.
For the implementation of the custodial protection
for $Z b \bar{b}$ coupling, the
required charges are:
\begin{itemize}

\item
$T_{ 3 R } = -1/2$ for $(t,b)_L$ and
thus $T_{ 3 R }
= 0, \, -1$ for $t_R$ and $b_R$
to obtain the top and bottom masses\footnote{In the model where
top and bottom masses are obtained from the same $5D$
$(t, b)_L$ multiplet.},
respectively.
\end{itemize}
Thus, we must modify the
Pati-Salam representations.
Moreover, the $SU(2)_L$ and $SU(2)_R$
$5D$ gauge couplings must be equal.

However,
the above requirement does not completely fix the model: we first discuss
the relevant parameters left over below and then describe a variety of
models with specific choices of these parameters.

\subsubsection{Composite charge leptons}
Once we resort to
non-canonical representations, we can
choose
\begin{itemize}
\item
$T_{ 3 R } = 0$ for $\tau_R$ (and other RH charged leptons) in order to provide custodial protection for its
coupling to $Z$ as well.
\end{itemize}
In this way,
$\tau_R$ can be localized very close to the TeV brane\footnote{In
order to obtain the charged lepton mass hierarchy,
$e_R$ and $\mu_R$ might have to be localized farther away from
the TeV brane than the $\tau_R$.},
i.e., we can contemplate larger couplings of KK $\tau_R$
to gauge KK modes (in particular, $Z^{ \prime }$).
Since,
via AdS/CFT correspondence, such a scenario is dual to
$\tau_R$ being a composite particle of $4D$ strong dynamics,
we will refer to this feature as ``composite'' $\tau_R$\footnote{
Note that the custodial symmetry cannot protect shift in
$Z$ coupling to LH charged leptons and LH neutrinos {\em simultaneously}
since we require $T_{ 3 R } = T_{ 3 L }$ for this purpose and LH charged
lepton and LH neutrino obviously have different $T_{ 3 L }$, but the same $T_{ 3 R }$.}.
Then we must choose
$T_{ 3 R } = +1/2$ for $( \nu, \tau )_L$ to obtain
charged lepton masses.
%

One may wonder whether this possibility of having a composite
$\tau_R$ is
constrained by precision tests. For instance, virtual KK
$Z$ boson exchange
will generate $4$-fermion operators involving
$\tau_R$ and other SM fermions.
In our case the dominant
constraint comes from
$\left(\bar{e} \gamma^{ \mu } e\right)\left( \bar{ \tau } \gamma_{ \mu } \tau_R\right)$ operator since the couplings of
KK $Z$ bosons to electrons are
vector-like in nature, whereas
the tau's are RH as discussed above.
Using the LEP bounds on such contact interactions from~\cite{:2003ih},
we find that the effective scale
suppressing this higher dimension
operator should be at least 3\,TeV.
In our case, the KK $Z$ coupling to $\tau_R$ is (roughly)
given by
$\sim g_Z  \sqrt{ k \pi  R }$ while the electron coupling
is $\sim g_Z / \sqrt{ k \pi  R }$
which gives roughly a coefficient of $1 / (  4 \rm\,TeV )^2$
for this operator for a $3\,$TeV KK mass scale,
and hence is consistent with the bounds.
However, with a composite $\tau_R$, constraints from lepton flavor violation might still be an issue
which can be addressed by gauging (at the $5D$ level) the flavor symmetries~\cite{Align}.

\subsubsection{DM couplings to $Z'$}
Of particular importance are obviously the representation and hence
coupling of the DM
candidate, $\nu'$. Since the couplings of $Z^{ \prime }$ are
in general of the same form
as canonical model (albeit with a different
$\sin \theta^{ \prime }$) and DM is a SM gauge singlet ($Y=0$),
its coupling to $Z^{ \prime }$ is proportional to
$T_{ 3 R }$, {\it i.e.},
the coupling of $\nu^{ \prime }$ to $Z^{ \prime }$ is vanishing
(non-vanishing)
for $T_{ 3 R }^{ \nu^{ \prime } } = 0$ ($\neq 0$).
In the case $T_{ 3 R }^{ \nu^{ \prime } }
= 0$, the DM coupling to the SM $Z$ (of course
induced by higher-order effects)
is also custodially protected~\cite{custodial2}.
Obviously,
the model's phenomenology differ qualitatively
depending on whether DM couples to $Z'$ (and $Z$) or not,
so that a crucial choice is
\begin{itemize}
 \item $T_{ 3 R }^{ \nu^{ \prime } } \neq 0$ vs. $T_{ 3 R }^{ \nu^{ \prime } } = 0$.
\end{itemize}
In the following parts we discuss two specific models
(two more are given in the appendix) which demonstrate
the essential differences.
Due to the fact that our DM is localized near the TeV brane
(just like other KK's), a
non vanishing  $T_{ 3 R }^{ \nu^{ \prime } } $
would imply  a sizable DM-$Z'$ coupling.
This case
tends to yield a too large an annihilation cross section via
$Z^{ \prime }$ exchange
into electroweak gauge bosons/top quarks
and hence typically
a too low relic density, unless the DM is of ${\cal O}(100)$ GeV in which case
direct detection from $Z$ exchange becomes a strong constraint.
Of course, if in addition $\tau_R$ is composite, then
the DM annihilation
into $\tau_R$'s
(which do couple to $Z^{ \prime }$) could be significant which could be interesting for indirect
cosmic ray positron/electron signal.

As we shall see later, it is quite remarkable that our model with the
smallest
fully unified, {\it i.e.}, $SO(10)$ representations,
actually predict
$T_{ 3 R }^{ \nu^{ \prime } } = 0$. Thus, leading to
vanishing coupling of DM to $Z^{ \prime }$
and SM $Z$, making it compatible with observed DM density and direct detection
bounds.

Finally,
in case where $T_{ 3 R }^{ \nu^{ \prime } } = 0$ ($\neq 0$) we also require
$X^{ \nu^{ \prime } } = 0$ ($\neq 0$)
in order to obtain $Y^{ \nu^{ \prime } } = 0$
(again, in general hypercharge is a combination of
$T_{ 3 R}$ and $X$, but with a different one than in canonical model).
%
%
%

\subsection{Model I (a): $T_{ 3 R } ^{ \nu^{ \prime } } \neq 0$
and custodial for leptons}

One possible choice of non-canonical Pati-Salam representations
satisfying the above conditions for
cosmic ray signals in positron/electron is given in table~\ref{custodial1}.
The SM
hypercharge is then given by
\begin{eqnarray}
Y & = & T_{ 3 R } + \sqrt{1/6} X\,,
\end{eqnarray}
and the DM and $t_R$ arise from a {\bf 35} of $SU(4)$.
The couplings to $Z^{ \prime }$ are then given by
\begin{eqnarray}
{ g_{ LR }  \over \cos \theta^{ \prime }
} \left( T_{ 3 R } - Y \sin^2 \theta^{ \prime }
\right)
\nonumber
\end{eqnarray}
as before, but with
$\sin^2 \theta^{ \prime }_{\mathbf{35}}
\equiv\left( 6  g_4^2 \right) / \left( 6  g_4^2 + g_{ LR }^2 \right),$ instead of the canonical value
due to the modified combination of $T_{ 3 R }$
and $X$ entering the hypercharge
(note that $g_R$ of before is replaced by
$g_{ LR }$ due to equality of $SU(2)_{ L, \; R }$ couplings).
 $\sin^2 \theta^{ \prime }$ is a free parameter
at the level of Pati-Salam gauge group.
We will leave a detailed analysis of completing this
model into $SO(10)$-type full unification, including
calculation of the resulting gauge coupling unification
in this model, for future work.
Here, we simply note a few
features of a potential unification into $SO(10)$.
First, such an extension
seems to require $SO(10)$ representations larger than $\bf{560}$~\cite{Slansky:1981yr}.
Moreover, even if we find such a representation, the
normalization of hypercharge above is
not the usual $SU(5)$ one so that this model
does not maintain even the SM-level of
unification of couplings.

However, a loop-level matching of the $5D$ gauge couplings
to the observed QCD and $SU(2)_L$ ones with the assumption of
small tree-level brane kinetic terms gives $g_{ LR } \approx g_4$
(just like the canonical $SO(10)$ case). Based on this observation,
we can choose $g_{ LR } \approx g_4$
({\it i.e.}, $\sin^2 \theta^{ \prime }
\approx 6/7$)
as
a ``benchmark'' value for this Pati-Salam model.
It is crucial to realize that the above model is just one choice satisfying
the conditions of custodial symmetry for the $Z b \bar{b}$ coupling
so that the value $T_{ 3 R } = -1$ (giving $Y = 0$) for $\nu^{ \prime }$
(and similarly the value of $\sin ^2 \theta^{ \prime }$, even with
the assumption of $g_{ LR } \approx g_4$)
is not unique: see the model below and the two models in
appendix~\ref{othercust}.


\begin{table}[hbt]
\begin{center}
\begin{tabular}{|c|c|c|c|}
\hline
& $SU(4)_c \sim SU(3)_C \times U(1)_X$ & $SU(2)_L$ & $SU(2)_R$ \\
\hline
$t_R$, $\nu^{ \prime }$ & \bf{35} $\sim$ \bf{3}$_{ \frac{8}{3} }$,
\bf{1}$_{4}$... & \bf{1} & \bf{3}
\\
\hline
$(t,b)_L$ & \bf{35} $\sim$ \bf{3}$_{ \frac{8}{3} }$,... & \bf{2} & \bf{2}
\\
\hline
$\tau_R$ & \bf{ $\bf{ \overline{35} }$ } $\sim$
\bf{1}$_{ - 4 }$,... & \bf{1} & \bf{1} \\
\hline
$(\nu, \tau )_L$
& \bf{ $\bf{ \overline{35} }$ }
$\sim$
\bf{1}$_{ - 4 }$,... & \bf{2} & \bf{2} \\
\hline
$b_R$
& \bf{35}
$\sim$
\bf{3}$_{ \frac{8}{3} }$,... & \bf{1} & \bf{3} \\
\hline
\hline
$H$ & \bf{1} & \bf{2} & \bf{2} \\
\hline
\end{tabular}
\end{center}

\caption{An example for a model with custodial
representations for $b_L$  and RH charged leptons, with
non-vanishing $\nu^{ \prime } \bar{ \nu^{ \prime } } Z^{ \prime }$
coupling (see Tab.~\ref{couplingscust1}),  the subscripts denote
the $ \sqrt{8/3} \; X$ charge.}
\label{custodial1}
\end{table}


\subsection{Model II: smallest full unification
%
%
}
We shall now construct a Pati-Salam model based on the {\bf 15} representation of SU(4) and
show that it is compatible with full unification
into $SO(10)$\footnote{See
also reference~\cite{gripaios}.}. The model also has
$SU(5)$ normalization for hypercharge, and
it predicts vanishing $T_{ 3 R }^{ \nu^{ \prime }}$
and hence DM coupling to $Z^{ \prime }/Z$.


The Pati-Salam
model is shown in table~\ref{custodialGUT}, it
can be fully unified into the following $SO(10)$
representations: ${\bf 45}$ for $t_R$ and $b_R$,
 ${\bf 120}$
for $(t,b)_L$ and the canonical, i.e., ${\bf 16}$ for leptons.
So, RH charged leptons are not protected by the custodial symmetry,
but the model can be modified easily to include this feature:
for example, LH and RH leptons being $( {\bf 10, 2, 2} )$
and $( {\bf 10, 1,1 })$
under $SU(4)_c \times SU(2)_L \times SU(2)_R$, respectively, which fit into
${\bf 210}$ and ${\bf 120}$, respectively
of $SO(10)$.
Moreover, the hypercharge normalization
\begin{eqnarray}
Y & = & T_{ 3 R } - \sqrt{ \frac{2}{3} } X\,.
\end{eqnarray}
is the {\em same} as in $SU(5)$ so that this model
maintains SM-level of
unification of couplings when fully unified into $SO(10)$.

\subsubsection{Vanishing coupling of $Z'$ to $\nu'$ pair ($T_{ 3 R }^{ \nu^{ \prime } }=0$)}

Note that
the
$X$-charge of $\nu^{ \prime }$ vanishes (see Tab.~\ref{custodialGUT})
for this choice of $t_R$ representation so that
the $\nu^{ \prime } \bar{ \nu^{ \prime } } Z^{ \prime }$ and
$\nu^{ \prime } \bar{ \nu^{ \prime } } Z$
couplings vanish.
Thus, this case might be uninteresting for indirect searches
for DM annihilation in cosmic ray positrons/electrons,
{\em irrespective} of custodial symmetry for RH charged leptons --
that is why we simply chose the canonical representations
for leptons in table~\ref{custodialGUT}.
However, as shown below, it
may lead to an observed future signal due to anomalously large antiproton flux in the hundreds of GeV region and has the benefit of
yielding a correct DM relic abundance.
And, with the custodial symmetry for RH leptons,
LHC signals related to composite $\tau_R$
become a possibility.

\begin{table}[hbt]
\begin{center}
\begin{tabular}{|c|c|c|c|}
\hline
& $SU(4)_c \sim SU(3)_C \times U(1)_X$ & $SU(2)_L$ & $SU(2)_R$ \\
\hline
$t_R$, $\nu^{ \prime }$ & \bf{15} $\sim$ \bf{3}$_{ \frac{-4}{3} }$,
\bf{1}$_{0}$... & \bf{1} & \bf{1}
\\
\hline
$(t,b)_L$ & \bf{15} $\sim$ \bf{3}$_{ \frac{-4}{3} }$,... & \bf{2} & \bf{2}
\\
\hline
$\tau_R$ & \bf{4} $\sim$
\bf{1}$_{ 1 }$,... & \bf{1} & \bf{2} \\
\hline
$(\nu, \tau )_L$
& \bf{4}
$\sim$
\bf{1}$_{ 1 }$,... & \bf{2} & \bf{1} \\
\hline
$b_R$
& \bf{15}
$\sim$
\bf{3}$_{ \frac{-4}{3} }$,... & \bf{1} & \bf{3} \\
\hline
\hline
$H$ & \bf{1} & \bf{2} & \bf{2} \\
\hline
\end{tabular}
\end{center}
\caption{
An example for a model with custodial
representations for $b_L$  which results in simplest
full unification. Charged leptons are not protected by the custodial symmetry and the $\nu^{ \prime } \bar{ \nu^{ \prime } } Z^{ \prime }$
coupling vanishes (see Tab.~\ref{couplingscustGUT}).
The subscripts denote
the $ \sqrt{8/3} \; X$ charge.}
\label{custodialGUT}
\end{table}

The couplings to $Z^{ \prime }$ are then given by
$\left( g_{ LR } / \cos \theta^{ \prime }
\right) \left( T_{ 3 R } - Y \sin^2 \theta^{ \prime }
\right)$
as before, but with
$\sin^2 \theta^{ \prime }_{\mathbf 15}
\equiv \left( {3\over 2}  g_4^2 \right) / \left( {3\over 2} g_4^2 + g_{ LR }^2 \right)$ as in the canonical case.

%


\begin{table}[hbt]

\begin{center}

\begin{tabular}{|c|c|}
\hline
SM & $t_R$, $(t,b)_L$, $\tau_R$, $\mu_R$, $W$, $Z$, $h$ \\
\hline
\end{tabular}

\vspace{0.5cm}

\begin{tabular}{|c|c|}
\hline
Non-SM & Comments (quantum numbers) \\
\hline
$\nu^{ \prime }$ & DM: exotic RH $\nu$ (SM singlet) with $B = 1/3$  \\
\hline
$\phi$ & radion (scalar with Higgs-like coupling to SM) \\
\hline
$Z^{ \prime }$ & extra/non-SM $U(1)$ in GUT \\
\hline
$X_s$ & leptoquark GUT gauge boson \\
\hline
\end{tabular}

\end{center}

\caption{Particle content relevant for DM (in-)direct detection.}
\label{content}
\end{table}


\begin{table}[hbt]
\begin{center}
\begin{tabular}{|c|c|c|}
\hline
Coupling & Value (in units of
$g_{ LR } \sqrt{ k \pi R }$) & Comments \\
\hline
$\overline{ \nu^{ \prime }_R } \gamma_{ \mu } Z^{ \prime \; \mu}
\nu^{ \prime }_R$ & $ - a_{ \nu^{ \prime }_R }
\cos^{-1} \theta^{ \prime }$ &
$a_{ \nu^{ \prime }_R }  \sim 1$ \\
\hline
$\overline{ \hat{ \nu^{ \prime }_R } } \gamma_{ \mu }
Z^{ \prime \; \mu} \hat{
\nu^{ \prime }_R }$ & $ - a_{ \hat{ \nu^{ \prime }_R } }
\cos^{-1} \theta^{ \prime }$ & $a_{ \hat{ \nu^{ \prime }_R } } \sim
\left( \frac{ m_{ \nu^{ \prime } } }{ M_{\rm KK}} \right)^2 $
\\
\hline \hline
$\overline{ t_R } \gamma_{ \mu } Z^{ \prime \; \mu } t_R$ & $ - \frac{2}{3}
a_{ t_R }
\cos^{-1} \theta^{ \prime } \sin^2 \theta^{ \prime }$
& $a_{ t_R }  \stackrel{<}{\sim} 1$ \\
\hline \hline
$\overline{ (t, b)_L } \gamma_{ \mu } Z^{ \prime \; \mu } (t, b)_L$ &
$a_{ (t,b)_L } \cos^{-1} \theta^{ \prime } \left( -\frac{1}{2}
- \frac{1}{6} \sin^2 \theta^{ \prime }
\right)$ & $a_{ (t,b)_L }
\stackrel{<}{\sim} 1$ \\
& & such that $\sqrt{ a_{ t_R } \; a_{ (t,b)_L } } \sim \frac{1}{ Y_{ KK } } \sim \frac{1}{7}$ \\
\hline \hline
$\overline{ ( \nu, \tau )_L } \gamma_{ \mu } Z^{ \prime \; \mu }
( \nu, \tau )_L$ & $a_{ ( \nu, \tau )_L }
\cos^{-1} \theta^{ \prime } \left( \frac{1}{2}
+ \frac{1}{2} \sin^2 \theta^{ \prime }
\right)$ &
$a_{ ( \nu, \tau )_L } \stackrel{<}{\sim} \frac{1}{10}$ \\
\hline \hline
$\overline{ \tau_R} \gamma_{ \mu } Z^{ \prime \; \mu } \tau_R $ & $a_{ \tau_R }
\cos^{-1} \theta^{ \prime } \sin^2 \theta^{ \prime }$ &
$a_{ \tau_R } \stackrel{<}{\sim} 1$ \\
\hline \hline
$\overline{ b_R } \gamma_{ \mu } Z^{ \prime \; \mu } b_R$ & $a_{ b_R }
\cos^{-1} \theta^{ \prime } \left( - 1
+ \frac{1}{3} \sin^2 \theta^{ \prime }
\right)$ &
$a_{ b_R } \stackrel{<}{\sim} \frac{1}{10}$ \\
\hline \hline
$Z_{ long. } Z^{ \prime }_{ \mu } h$ & $a_{ Z^{ \prime } H } \frac{ \cos \theta^{ \prime } }{2}
\left(  p^{ \mu }_{ Z_{ long. }} - p^{ \mu }_{h}
\right)$
& $a_{ Z^{ \prime } H } \sim 1$ \\
\hline
$W^+_{ long. } Z^{ \prime }_{ \mu } W^-_{ long. }$ &
$a_{ Z^{ \prime } H } \frac{ \cos \theta^{ \prime } }{2}
\left( p^{ \mu }_{ W^+_{ long. } } - p^{ \mu }_{ W^-_{ long. } } \right)$
& $a_{ Z^{ \prime } H } \sim 1$ \\
\hline
\hline
$\overline{ \nu^{ \prime }_R } \hat{ \nu^{ \prime }_R } \phi$ (radion) &
$\frac{ m_{ \nu^{ \prime }_R } }{ \Lambda_r }$ (no
$g_{ LR } \sqrt{ k \pi R }$) & $ \Lambda_r \equiv
\sqrt{6} M_{ Pl. } e^{ - k \pi R } $ \\
\hline
%
%
\end{tabular}
\end{center}

\caption{Couplings relevant for DM annihilation in model
with custodial
symmetry for $Z b \bar{b}$
 and RH charged leptons, with
non-vanishing $\nu^{ \prime } \bar{ \nu^{ \prime } } Z^{ \prime }$
coupling (see Tab.~\ref{custodial1}):
value of $\sin^2 \theta^{ \prime }$ is $6/7$ and note that $T_{ 3 R }^{ \nu^{ \prime } }=1$.
}
\label{couplingscust1}
\end{table}

\begin{table}[hbt]
\begin{center}
\begin{tabular}{|c|c|c|}
\hline
Coupling  &Value &Comments \\
\hline
$\overline{ \nu^{ \prime }_R } \gamma_{ \mu } X_s^{ \mu }
t_R$ & $ \sqrt{k\pi R}\, { g_4 \over \sqrt2}  a_{ t_R \nu^{ \prime }_R } $ &
$a_{ t_R \nu^{ \prime }_R } \sim \sqrt{ a_{ t_R } }$ \\
\hline \hline
$\overline{ \nu^{ \prime }_R } \hat{ \nu^{ \prime }_R } \phi$  &
$\frac{ m_{ \nu^{ \prime }_R } }{ \Lambda_r }$ & same as in Tab.~\ref{couplingscust1} \\
\hline
%
%
\end{tabular}
\end{center}

\caption{Couplings relevant for DM annihilation in simplest
fully unifiable custodial case (see Tab.~\ref{custodialGUT}):
value of $\sin^2 \theta^{ \prime }$ is $3/5$, but largely irrelevant for cosmology since $T_{ 3 R }^{ \nu^{ \prime } }=0$.
}
\label{couplingscustGUT}
\end{table}

\subsection{Summary of models characterization}

The
relevant particle content and their couplings
are summarized in tables \ref{content} and \ref{couplingscust1} (\ref{couplingscustGUT}) for the partial (fully) unifiable models respectively.
A few
comments about the tables are in order:

\begin{itemize}

\item
$\nu^{ \prime }$ is the SM singlet (i.e.,
with quantum numbers of a RH neutrino) GUT partner
of $t_R$\footnote{Since, with custodial protection
of $Z b \bar{b}$ coupling, $(t,b)_L$
can also be close to the TeV brane, it is possible that
the LZP comes from this multiplet instead of
$t_R$. The analysis for the two cases is
similar.}, but with (exotic) baryon number of $1/3$.
$\nu^{ \prime }_R$ denotes its RH chirality
and has a profile localized near the TeV brane (like for any other
KK mode), irrespective of
bulk mass ($c$) parameter for this GUT multiplet
\footnote{We neglect any GUT breaking here in the $5D$
fermion mass parameters within a GUT multiplet unlike references
\cite{Agashe:2004ci}
where small splittings of this type were allowed.}
which dictates the profile of $t_R$.

\item
Following the notation of
references~\cite{Agashe:2004ci},
$\hat{ \nu^{ \prime }_R }$ denotes the Dirac partner ({\em left}-handed) of
$\nu^{ \prime }_R$\footnote{$\nu^{ \prime }_L$ was used in references
\cite{Agashe:2004ci} for $SU(2)_L$ doublet from $(t,b)_L$ multiplet.}. Its
profile does depend on $c$ for $t_R$ in such a way that it moves
farther away from the TeV brane
as $t_R$ gets closer to the TeV brane --
the $\nu^{ \prime }$ mass ($\propto$ this
overlap) decreases in this process.

\item
$X_s$ (mostly relevant for the unifiable model with no DM-$Z'$ coupling) and $Z^{ \prime }$ (relevant for the partially unified model where DM-$Z'$ coupling
controls the resulting relic density) are, respectively, the non-abelian and
$U(1)$ gauge bosons (beyond gluons)
contained in $SU(4)_c$ and have masses (almost) same as
those of KK modes of SM gauge bosons (denoted by $M_{\rm KK }$).

\item
Neglecting TeV brane-localized kinetic terms for gauge fields,
the couplings can be conveniently expressed (as in the middle column of tables~\ref{couplingscust1} and~\ref{couplingscustGUT} )
in units of $ g_{ 4D } \sqrt{ k \pi R } \equiv g_{ 5D } \sqrt{k}$,
where $g_{ 5D }$ is the $5D$ gauge coupling (of mass dimension $-1/2$)
such that $g_{ 4D }$ is the coupling of the (``would-be'' in some cases)
zero-mode (and hence is volume suppressed compared to $g_{ 5D }$).

\item
The custodial symmetry for $Z$ couplings to fermions
requires the two $SU(2)$ $5D$ couplings to be equal, but
the $SU(4)_C$ coupling is unrelated to it. Hence, there appear
two $g_{ 4D }$'s in the table: $g_{ LR }$
for the two $SU(2)$ groups and
$g_4$ for $SU(4)$ group.

\item
$g_{ 4D }$'s cannot always be {\em equated} to the
SM gauge couplings since the relation between
the two couplings depends
on presence of tree-level UV brane kinetic terms and
also loop corrections.
A detailed analysis is left for future work, but here
we can choose each of the $g_{ 4D }$'s to be (independently)
roughly between the SM hypercharge and QCD couplings,
{\it i.e.}, $0.35 \stackrel{<}{\sim} g_{ LR }, \; g_4 \stackrel{<}{\sim} 1$.

\item
The factors $a$'s in middle column of tables \ref{couplingscust1} and \ref{couplingscustGUT}  come from
overlap of wavefunctions in the extra dimension of the involved modes.
Specifically, for a coupling of
3 (usual) KK modes (which are localized near
TeV brane), the overlap gives $a \sim 1$.
Then, each time we replace a KK mode by SM/zero-mode we
incur a ``cost'' of $\sqrt{a_{ SM }}$ which is
roughly the {\em ratio} of profile of
SM fermion/zero-mode at/near the TeV brane to that of a KK fermion (or
equivalently
the degree of ``compositeness'' of these SM fermions in the dual
CFT description).

\item
Similarly,
$\sqrt{ a_{ \hat{ \nu^{ \prime }_R } } }$ is the degree of compositeness of
$\hat{ \nu^{ \prime }_R }$, i.e., the
ratio of its profile near the TeV brane to that
of a usual
KK fermion (which is localized near TeV brane). With $\nu^{ \prime }_R$
being fully composite
(i.e., localized near the TeV brane), the particular appearance
of $\sqrt{ a_{ \hat{ \nu^{ \prime }_R } } }$ in the table
is thus explained.

\item
We require $\sqrt{ a_{ t_R } \; a_{ (t,b)_L } }
\sim 1 / Y_{ KK }$
such that we can obtain top Yukawa -- given by $Y_{ KK }
\sqrt{ a_{ t_R } \; a_{ (t,b)_L } }$ -- of $1$: here, $Y_{ KK }$ is the
coupling of 2 KK fermions to Higgs and we require it to be
smaller than $\sim 1/7$ to allow $\sim 3$ KK modes
in the $5D$ effective field theory

\item
The mixing angle $\sin^2 \theta^{ \prime }
\equiv \left( 6  g_4^2 \right) / \left( 6 g_4^2 + g_{ LR }^2 \right)$ appearing in $Z^{ \prime }$
couplings is a
free parameter (since $g_{ LR }$ is
unrelated to $g_4$), but a
``benchmark'' value for this mixing angle is $6/7$.

\item
We use
equivalence theorem so that $W/Z_{ long. }$ is the unphysical Higgs.

\item
Finally, the coupling of $\nu^{ \prime }$ to radion has an additional
dependence on $c$ for $t_R$ only for the case $m_{ \nu^{ \prime } }
\stackrel{<}{\sim} M_{\rm KK } / \sqrt{ k \pi R }$ which occurs
for $c$ for $t_R \stackrel{<}{\sim} -1/2$ (in the convention that
$c = 1/2$ is flat profile for $t_R$). Since we are most likely
not interested in this DM mass region, no factor of $a$ is shown here
in the coupling of DM to radion.

\end{itemize}

\section{Implications for Cosmology and Astrophysics}
\label{annihilation}

The potentially light radion, an intrinsic ingredient of the model, has significant implications for cosmology and astrophysics.
The existence of a light degree of freedom opens the possibility of an enhancement factor in the velocity weighted annihilation
cross section, relevant for the current epoch, compared to the cosmological value at freeze-out. This effect occurs via the Sommerfeld Enhancement (SE)~\cite{Somm}.
An enhancement is required in order for annihilation signals to overcome astrophysical backgrounds, which would drown those
signals for a TeV thermal relic with canonical cross section $\langle\sigma v\rangle\sim\mathrm{ \ few \ }10^{-26}{\,\rm cm}^3{\rm \, s}^{-1}$.

In section \ref{ssec:SE} we explore the SE arising in our framework. Requiring a very large enhancement dictates
special correlations between model parameters, as well as constrains the radion mass. In section \ref{ssec:relicdd} we proceed
to identify the parameter region compatible with direct detection limits and with the DM relic density implied by WMAP data. We find that a sizable SE factor is possible, and that 
the model consistent with full unification is viable over a large region of parameter space.

Indirect detection searches in Galactic cosmic rays, including high energy gamma rays and neutrinos as
well as antiprotons, provide constraints on the viable magnitude of the SE factor. We study those limits in sections \ref{sssec:phot} and \ref{sssec:pbar}.
For antiproton energies $\epsilon\gsim10$\,GeV, no detailed assumptions are required regarding the propagation in the Galaxy. We study possible imprints of our model in the high energy antiproton flux, accessible to existing and near future experiments. We find that in a sizable fraction of our parameter space (with heavy DM and a PGB radion) a $\bar p/p$ future signal is quite generic.

Regarding CR positrons, as discussed in the introduction, an intriguing hint was reported by the PAMELA experiment, suggesting a spectral behavior which can not be easily reconciled with simple diffusion models of CR propagation~\cite{Adriani:2008zr}. To our view, this latter observation does not necessitate an exotic injection mechanism for the positrons, and we dedicate Sec.~\ref{sssec:pos} for a discussion of this point. Here we comment that our benchmark models which survive the requirements from direct detection, provide the correct DM relic density, and adhere with collider and precision test constraints, do not exhibit a large enough leptonic vs hadronic branching ratio as required to explain the positron fraction rise within the commonly adopted diffusion models.

\subsection{Sommerfeld enhancement with a light radion}\label{ssec:SE}

In this section we review the computation of the Sommerfeld enhancement factor, relevant for our
framework if the radion is much lighter than the dark matter particle~\cite{Somm}.
Requiring the maximal level of enhancement, SE\,$\gsim10^4$, implies particular correlations between model parameters.
We outline these correlations and show, in addition, that lower values of SE\,$\sim10^2-10^3$ are easily accessible.

The Sommerfeld enhancement due to Yukawa interaction is found by solving the ODE~\cite{ArkaniHamed:2008qn}
\beq\label{eq:SE1} \left[\frac{d^2}{dx^2}+\frac{e^{-\epsilon_\phi
x}}{x}+\epsilon_v^2\right]\chi(x)=0\,,\eeq
with
\beq\label{eq:SEpar}\epsilon_v=\frac{v}{\alpha} \ , \ \
\epsilon_\phi=\frac{m_r}{\alpha M} \,, \ \
\alpha=\frac{\lambda^2}{4\pi} \eeq
and with the boundary conditions
\beq \chi(x\to0)\to0 \,, \ \
\chi(x\to\infty)\to\sin(\epsilon_vx+\delta)\,.\eeq
Above, $M$ is the mass of the annihilating particles, $v$ is the velocity of each particle in the center of mass (CM) frame, $\lambda$ is the Yukawa coupling and $m_r$ is the radion mass. The
enhancement factor is then given by
\beq {\rm SE}=\left|\frac{\frac{d\chi}{dx}(x\to0)}{\epsilon_v}\right|^2.\eeq

Using (\ref{eq:SEpar}) we find, for our model,
\beqa\label{eq:ephiev}\begin{split}&\lambda=\frac{M}{\Lambda_r} \,, \ \
\alpha=7.9\cdot10^{-2}\left(\frac{M}{\Lambda_r}\right)^{2}\,,\\
&\epsilon_v=6.3\cdot10^{-3}\left(\frac{v}{150\,\rm km\,s^{-1}}\right)\left(\frac{M}{\Lambda_r}\right)^{-2}\,,\\
&\epsilon_\phi=1.2\cdot10^{-1}\left(\frac{m_r/M}{10^{-2}}\right)\left(\frac{M}{\Lambda_r}\right)^{-2}\,.\end{split}\eeqa
To get the effective enhancement one needs to average the SE over the DM velocity distribution, which we take as Maxwell-Boltzmann:
\beq f(v)\propto v^2e^{-v^2/2\sigma^2}\,,\eeq
where $\sigma$ is the rms velocity, $\sigma=\sqrt{\int
dvf(v)v^2/3}$\,.
Here we use $\sigma=\rm 150\,{\rm kms}^{-1}$
~\cite{Belli:2002yt,Governato:2006cq}. Uncertainties of $\mathcal{O}(1)$ associated with the value of the DM velocity distribution could modify some of the details of our results, notably when a maximal level of enhancement is considered; yet they would not change the overall conclusions nor the detailed results in cases where only moderate enhancement levels of order SE\,$\sim10^2$ are discussed.

\begin{figure}[hbp]\begin{center}
\hspace{-2cm}$\begin{array}{lcr}
\includegraphics[width=90mm,height=90mm]{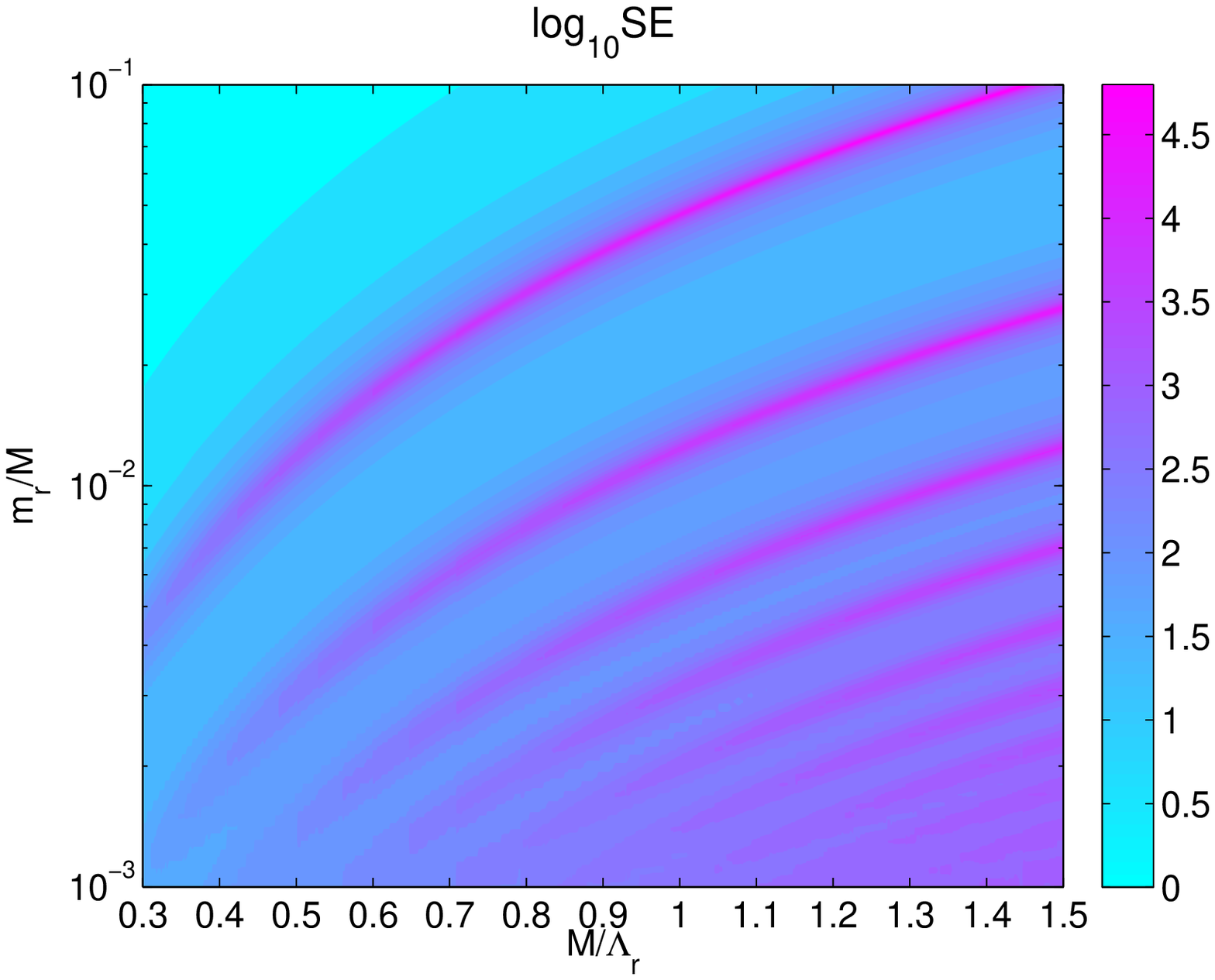}  &
\includegraphics[width=90mm,height=90mm]{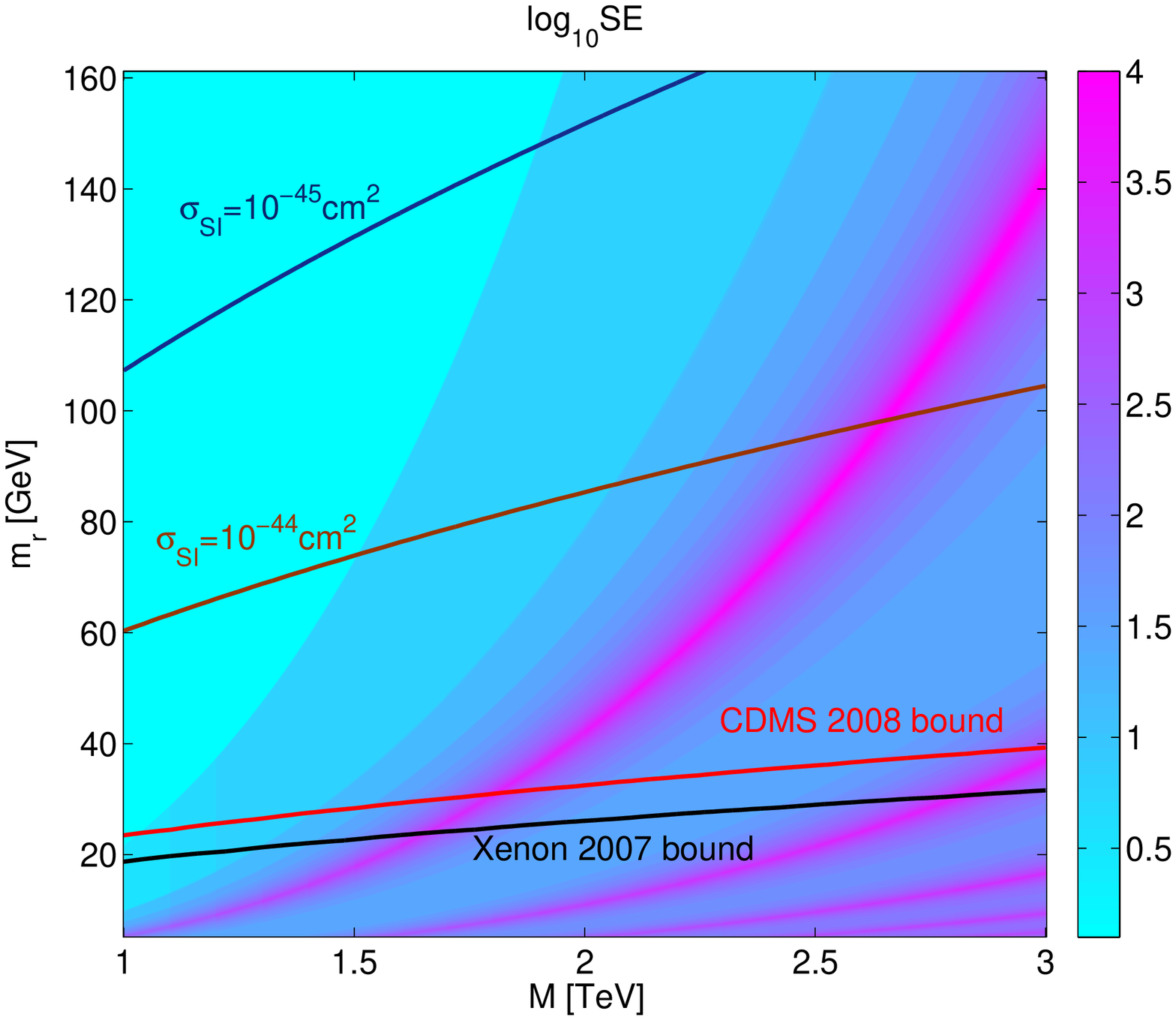}
\end{array}$
\caption{Left: SE factor projected onto the ($m_r/M,M/\Lambda_r$) plane, $m_r\,,M\,,\Lambda_r$
stand for the radion mass, the dark matter mass and the scale which suppresses the radion's couplings. Right: SE and
direct detection bounds, projected onto the $(m_r\,,M)$ plane at fixed $\Lambda_r=3$\,TeV.} \label{fig:DD_SE_Mdm_Mr_KK3}
\end{center}
\end{figure}

The Sommerfeld factor attainable for our model is depicted in Fig.~\ref{fig:DD_SE_Mdm_Mr_KK3}. In the left panel, the SE is plotted in the
$(M/\Lambda_r\,,m_r/M)$ plane. Resonance branches cross the $(M/\Lambda_r\,,m_r/M)$ plane, with enhancement factor SE\,$\gsim10^4$ attainable at the peak of each branch and values of SE\,$\sim10^3$ at peak vicinity. The location of the i'th resonance branch in the $(M/\Lambda_r\,,m_r/M)$ plane follows contours of constant values of $\epsilon_\phi=\epsilon_{\phi,i}$\,, with $\epsilon_{\phi,i}\approx0.6,0.15,0.07,...$ arising in the numerical solution to the Yukawa problem. Using Eq.~(\ref{eq:ephiev}) we see that the resonance branches correspond to parabolas,
\beq\frac{m_r}{M}\approx C_i\left(\frac{M}{\Lambda_r}\right)^2\,,\eeq
where the $C_i$s are constant numbers. Sample values are $C_1\approx0.05, \ C_2\approx0.01$ for the first (upper) two resonance branches.
We see that in order to obtain SE\,$>10^3$\,, significant correlation is required between $m_r$\,, $M$ and $\Lambda_r$\,.
Below we exploit this correlation to extract benchmark model points with interesting consequences for indirect
signatures in galactic cosmic rays. However, since in our case no large SE is required the
parameters need not be tuned, as to lie on the resonances in the effective potential, hence
the sensitivity to small changes in the relevant parameter is probably only polynomial.
The benchmark models we will consider can easily be located on the right panel of Figure~ \ref{fig:DD_SE_Mdm_Mr_KK3}\,, in which we plot the SE in the $(m_r\,,M)$ plane for a fixed value of $\Lambda_r=3$\,TeV. Direct detection constraints (discussed in the next section) are also superimposed on this panel.

Finally, note that the constraint discussed in reference~\cite{Feng:2009hw}
from correlation between Sommerfeld enhancement and relic density
is {\em not} applicable in our case since the relevant particles involved
in the two processes are different.
Furthermore, since our force carrier, the radion, is not ultra-light, higher partial waves beyond $s$-wave
are negligible in the current epoch annihilation, and constraints due to
enhanced DM self-scattering~\cite{Feng:2009hw,Buckley:2009in} are easily satisfied.

\subsection{Dark matter relic density and direct detection limits}\label{ssec:relicdd}

\subsubsection{Relic density}
As already anticipated the DM abundance is correlated with the DM-$Z'$ coupling size, in particular whether $T_{ 3 R }^{ \nu^{ \prime } }$ vanishes or not,
we discuss the two cases separately in the following.
The analytical expressions for the various annihilation cross section can be found in~\cite{Agashe:2004ci}, here we only
discuss the main qualitative feature of the models parameter space.
We compute the annihilation cross-section
using micrOMEGAs 2.2~\cite{Belanger:2006is} for the numerical evaluation of the freeze--out DM abundance (for simplicity we have set the KK $Z-Z'$ mixing to zero~\cite{Agashe:2004ci,Agashe:2007ki}).

One important feature of our models is that our DM candidate mass,
being the RH top partner, is correlated with the
localization of (or in $4D$ dual language, the amount
of compositeness of) $t_R$, which in turns
controls the relic density~\cite{Agashe:2004ci}:
\beq m_{\rm DM}(c) \approx \left\{ \begin{array}{ll}
0.65 \,(c + 1) \,M_{\rm KK} & \textrm{if $c>-0.25$}\\
0.83 \,\sqrt{c + {1\over2}}\, M_{\rm KK} & \textrm{if $-0.25>c>-0.5$}\\
0.83 \,\sqrt{c^2 - {1\over4}} \, M_{\rm KK} \exp{\left[k\pi R\left(c+ {1\over2}\right)\right]}  & \textrm{if $c<-0.5$}\\
\end{array} \right.\label{MDMc}
 \eeq
where $M_{\rm KK} \approx 2.5 \tilde{k}$
(with $\tilde k=k\exp{\left(-k\pi R\right)}$)
is the leading order $(++)$ KK gauge boson mass
and $c$ stands for the $t_R$ bulk mass.
For instance for $c>-{1/4}$ one finds $m_{\nu'} \approx \tilde k \pi( 1+ c)/2$ and for $-0.4<c<-{1/4}$, $m_{\nu'} \approx 2\tilde k \sqrt{ 1/2+ c}$.

In our calculations, we have neglected (for simplicity)
brane localized kinetic term
(BKT) for bulk fields.
BKT's can in principle be used to control the total annihilation cross section
and direct detection rate as follows~\cite{BKTs}.
First, BKT's for
gauge fields
tend to lower the coupling of the lightest gauge KK modes to other particles localized near the
TeV brane.
Such BKT's also lower the gauge KK mass relative to the
going rate, $\tilde{k}$,
mentioned above.
However, electroweak precision tests (in particular, the $S$
parameter) put a lower bound of a few TeV on the lightest KK scale which is (roughly)
independent of the
coupling of this KK mode.
Combining these two features,
we see that the
annihilation cross-section and similarly direct detection can be
reduced by BKT's for gauge fields.
However,
$\tilde{k}$ is then larger than a few TeV which might
introduce a severe little hierarchy problem into the model.
In addition,
BKT's for fermions can modify the correlation
between DM mass and localization of $t_R$ and, in turn,
some of our conclusions.

We find that DM annihilation into two radions requires the DM pair (fermion-antifermion) to be in a
$p$-wave, and hence is suppressed (see also reference~\cite{Bai:2009ms})\footnote{In general,
if the interactions respect parity, then
only $p$-wave annihilation of fermion-antifermion into pair of identical scalars is 
allowed~\cite{Gopalakrishna:2009yz}.}.
Thus this channel is not relevant for calculation of relic density nor does it get SE. 

\paragraph{Non vanishing DM-$Z'$ coupling.}
For $T_{ 3 R }^{ \nu^{ \prime } }\neq0$, we find quite generally that
there is a tension between obtaining the correct relic density and
being consistent with bounds from direct detection.
%
%
This is associated with the large $\nu' \bar \nu' Z'$ coupling,
being enhanced relative to the SM gauge couplings by the RS volume
factor, $\sqrt{k\pi R}\sim6$ (or as expected via the AdS/CFT correspondence, being an inter-composite
coupling)\footnote{A smaller volume factor would thus result in suppression of the DM annihilation and direct detection
rates. For example, since the focus here is on unification, one can assume that the UV brane scale is actually the unification scale, instead of the canonical choice of Planck scale
which gives
$\sqrt{k\pi R}\sim6$.
However, we have verified that, since the unification scale is only two orders of magnitude below the
Planck scale, the resulting improvement is only incremental. Hence the conclusion
about viability of these models is basically unchanged.
One can in principle consider a {\em much} smaller RS volume~\cite{Agashe:2007jb,LRS}.
However, then the SM level of unification of gauge couplings
and hence the motivation for considering a GUT model (and, in turn, the above DM candidate) is lost.}.
Furthermore, the large $T_{ 3 R }^{ \nu^{ \prime }}$ ($=2$)
for ${\bf 10}$ of $SU(4)$
and
the fact that $\cos^2\theta'$ is significantly smaller than one
(see Tab.~\ref{couplingscust1}),
$\cos^2\theta'_{\mathbf{35}}\sim{1\over 7}$, for model with ${\bf 35}$ of
$SU(4)$ make the rates even larger..
Depending on the mass of $\nu'$ compared to the intermediate particle,
$\nu'\bar{\nu'}$ annihilate into
the SM particles dominantly through either s-channel (for $2M_{\nu'} \le M_{Z'}$) or t-channel annihilation (for $2M_{\nu'} > M_{Z'}$).
We find that in the former case the $Z'$ becomes broad enough such that resonance enhancement of the cross section strongly suppresses the relic abundance, for $M_{Z'}\sim 2M_{\nu'} $. For $M_{\nu'} \ll M_{Z'}$ the off-resonance cross section is suppressed by $M_{\nu'}^2/ M_{Z'}^2$.
Hence, for $M_{\nu'}\lesssim M_{Z'}/5$ the resulting density is in the right ball park.
The case with more massive DM, $2M_{\nu'}> M_{Z'}$ has no kinematical suppression factors and yield a negligible freeze--out
density.

\begin{figure}[hbp]\begin{center}
\hspace{-2cm}
\includegraphics[width=80mm]{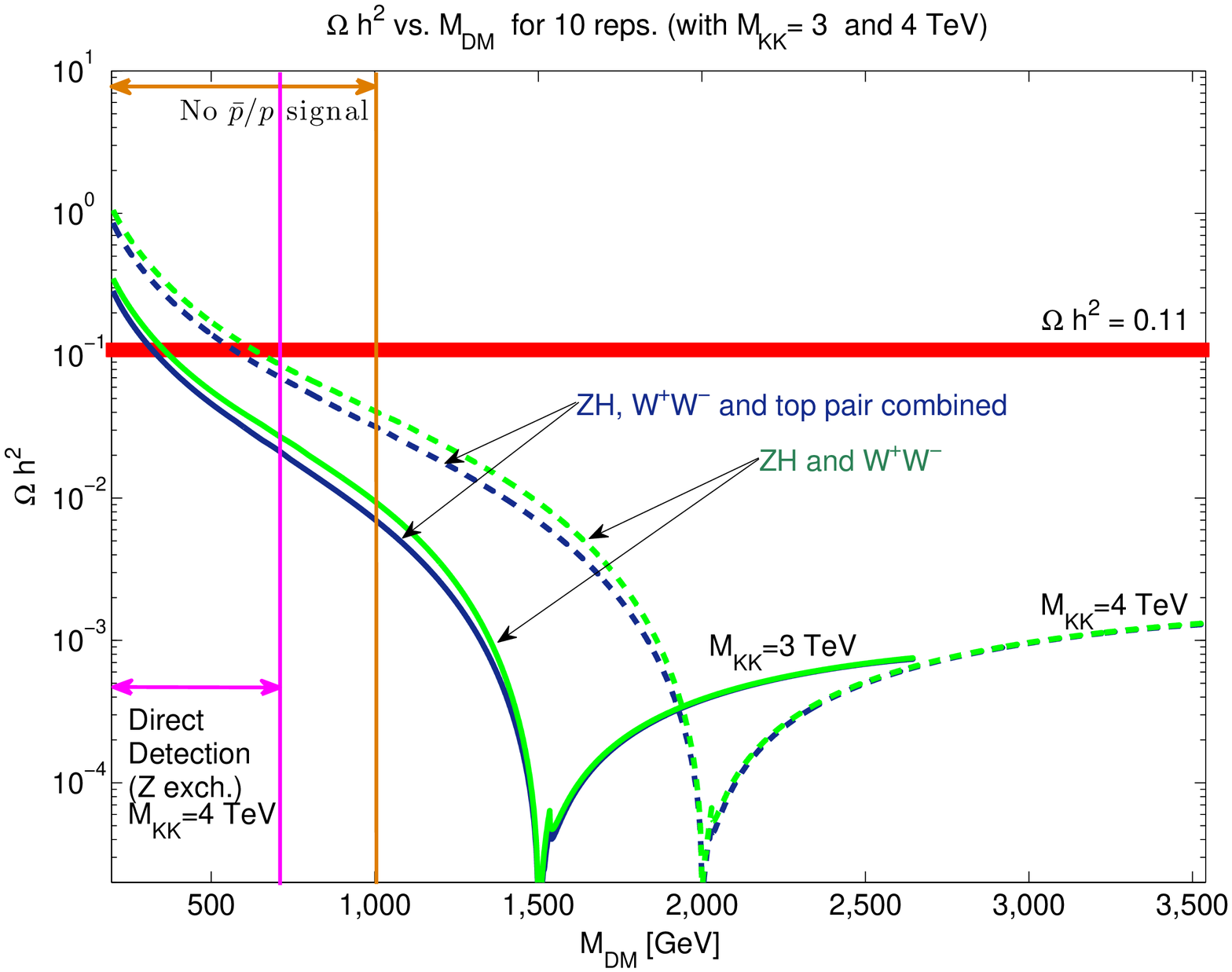}
\includegraphics[width=80mm]{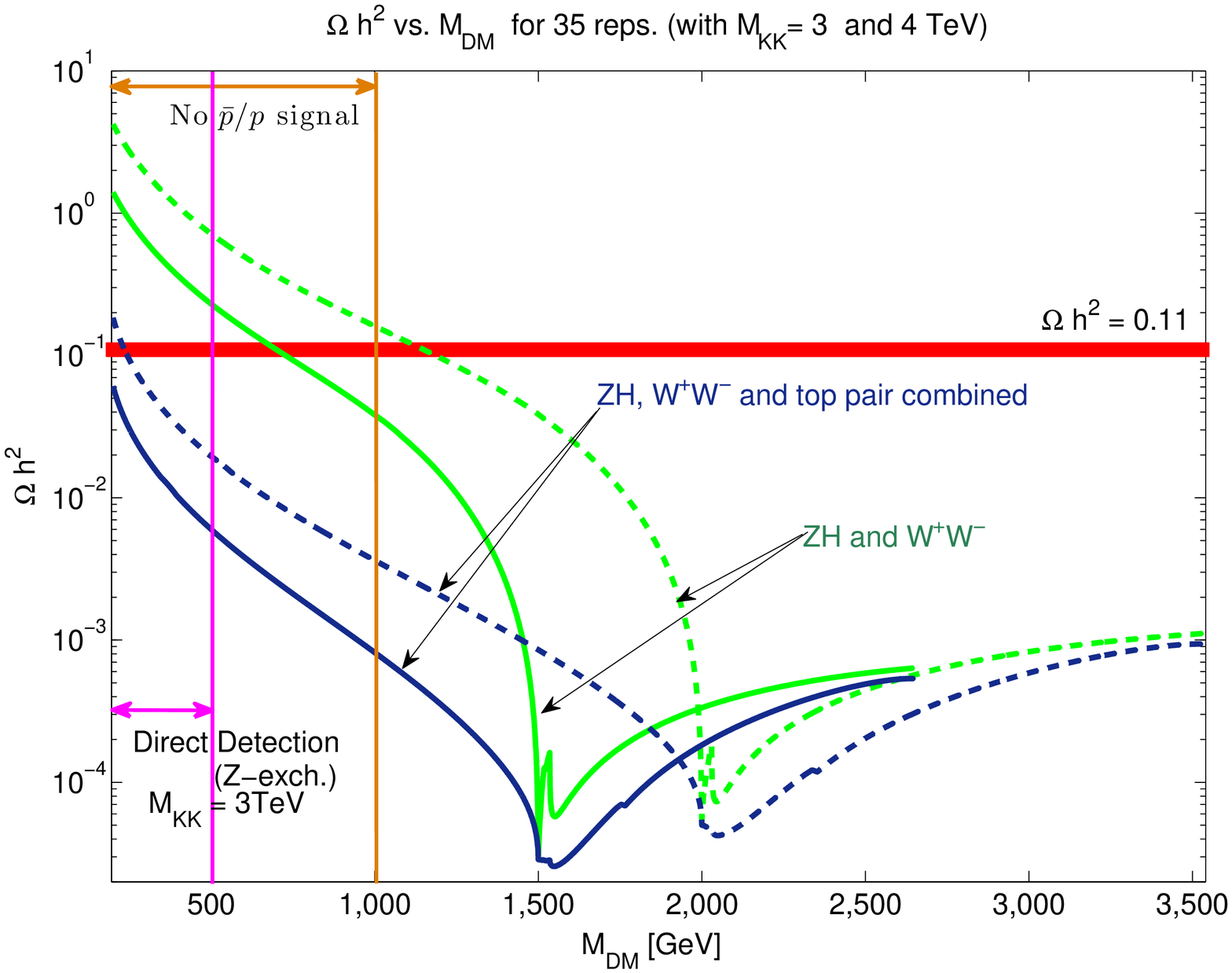}
\caption{Relic density, $\Omega h^2$ vs. the DM mass for the partially unified models with $\nu'\in\,${\bf 10}\,({\bf 35})
on the LHS (RHS). \label{fig:relic_density_Mdm}
Solid curves correspond to
$M_{\rm KK}=3\,$TeV and dashed ones to $M_{\rm KK}=4\,$TeV and $g_{LR}=g_{4}=0.35$.
Also shown, as vertical lines, are the constraints from direct detection (purple, left most vertical line)
and the region (gold) where typically no future $\bar p/p$ can be observed. Direct detection bound for $M_{\rm KK}= 3$\,TeV with {\bf 10} of $SU(4)$ is not shown
because for this case the entire range of DM mass
considered here is ruled out by the central value of the direct detection bound,
while $M_{\rm KK} = 4$\,TeV with {\bf 35} of SU(4) is not shown
because, in that case, the central value is $\sim 40$ GeV which is
below the smallest DM mass shown in the plot, {\it i.e.},
direct detection is a weak constraint in this case.}
\end{center}
\end{figure}

We show in Fig.~\ref{fig:relic_density_Mdm} the resulting ${\Omega_{\rm DM}} h^2$
 for partially unified models [$\nu'\in${\bf 10} as in Tab.~\ref{10} ({\bf 35}, Tab.~\ref{custodial1})]
 as a function of the DM mass.
 Curves are shown for
%
%
$M_{\rm KK }=3,4\,$ TeV, where
the green curve indicates the corresponding relic density only due to annihilation into the EW sector which is rather robust,
while the blue curve shows how the density is further suppressed when the coupling of $Z'$ to top quark pair is added (we used the canonical choice of $\sin\theta'$ given by setting $g_{LR}=g_{4}$, which is less robust).
The annihilation rate is calculated assuming the
%
%
mass relation of Eq.~(\ref{MDMc}), and the smallest possible coupling $g_{LR}=0.35$. This
choice of $g_{ L R }$ minimizes the rate, i.e., the resulting relic densities can be made much smaller by
allowing larger $g_{ LR }$, but not much bigger (which,
as we will show below, induces a strong constraint on these models).
%
%
%
Other parameters were not varied for concreteness, $M_{Z'}=M_{\rm KK}$,
$\Lambda_r=M_{\rm KK}$, and $T_{3R}^{\nu'}=1$ for relic density calculation.
%
%
%
For the model with ${\bf 10}$ of $SU(4)$, for $M_{\rm KK} =4$\,TeV, we see that there is a sizable
region of DM mass, {\it i.e.}, below $600$ GeV,
which gives correct relic density. On the other
hand, only the small region below $200$ GeV works for
model with ${\bf 35}$ of $SU(4)$.
However, as discussed in the following, typically both these regions implies
a too large rate for the direct detection experiments.

\begin{figure}[htp]\begin{center}
\hspace{-2cm}
\includegraphics[width=100mm]{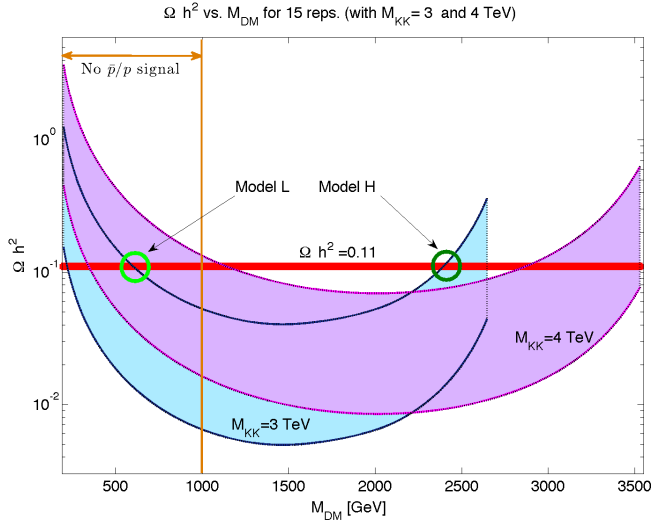}
\caption{Relic density, $\Omega h^2$ vs. the DM mass for the simplest fully
unifiable model with $\nu'\in$\,{\bf 15}. The bands correspond to varying the coupling in the favored range $g_{LR}=0.35-1$ and $M_{\rm KK}=3,4\,$TeV. The points relevant to our two bench mark models (``model L" and ``model H") are shown
as the two light and dark green circles respectively.
\label{fig:relic_density_Mdm_GUT}}
\end{center}
\end{figure}

\paragraph{Vanishing DM-$Z'$ coupling.}
In this case,
since DM coupling to $Z^{ \prime }$
vanishes, the dominant annihilation channel is
via $t$-channel $X_s$ exchange channel into final state heavy quarks, say
$t_R \bar t_R$ (see Tab.~\ref{couplingscustGUT}).
 As mentioned, the rate is controlled by the amount of compositeness of the RH tops which is also correlated (modulo
BKT's for fermions) with the DM mass as in Eq.~(\ref{MDMc}).
Thus, within this case an interesting correlation between
the DM mass and the resulting relic abundance is obtained.
This is also interesting in the context of precision GUT which probably requires composite RH top~\cite{Agashe:2005vg}, however, the issue of precision custodial unification is beyond the scope of this project.
We show in Fig.~\ref{fig:relic_density_Mdm_GUT} the resulting ${\Omega_{\rm DM}} h^2$
for the simplest fully unifiable model as a function of the DM mass.
Bands for $M_{Z'}=3,4\,$TeV are shown where for each $g_{LR}$ is scanned over the range $g_{LR}=0.35..1$, both within the favored range, other parameters were not varied for concreteness.
We see that there is a significantly larger region of the parameter space
(than in the previous models), {\it i.e.},
a few $100$ GeV to a few TeV, which yields
the correct DM abundance.
This feature is due mainly to absence of $Z^{ \prime }$ exchange
in the
simplest fully unifiable model.
However, the more important
impact of the absence of DM coupling to
$Z^{ \prime }$ is that this model is easily
consistent with bounds from direct detection
experiments (cf. other two models).

%
%


\subsubsection{Direct detection limits}
Many experiments are underway currently to directly detect dark matter, and still more are proposed to improve
the sensitivity. In order to ascertain the prospects of directly observing $\nu'$ in the model framework we are
considering, we compute the elastic $\nu'$-nucleon cross-section due to the $t$-channel exchange
of the radion, which is the most important channel, when radion mass is very light. The other important channel
is $t$-channel exchange of the $Z$, which was computed in~\cite{Agashe:2004ci, Belanger:2007dx}.

While contributions from radion exchange are generic within our framework
the ones
induced by $Z$ exchange (via $Z-Z'$ or via $\nu'_R$-KK $\nu^c_L$ mixing) only occur in the
partially unified model and not in the simplest fully unifiable unifiable model
(where the $\nu'$ coupling is custodially protected).

The cross-section for $Z$ exchange is
(roughly) independent of DM mass, but scales as $M_{\rm KK}^{-4}$. However,
the CDMS bound scales as $ 1 / M_{\rm DM }$
(i.e., inverse of number density of DM)
hence becomes dominant at low masses
and
%
%
tightly constrains the light DM region as shown by the (left-most) purple vertical lines of
Fig.~\ref{fig:relic_density_Mdm}. Several astrophysical unknowns (such as the local DM profile, the velocity
distribution etc.) are involved in converting the direct detection bound into a constraint on a microscopical
model parameter space (see {\it e.g.}~\cite{Bottino:2005qj}). Nevertheless, for concreteness,  taking central
values seriously, only a very small region survives for $M_{\rm KK } = 4$\,TeV (none for $M_{\rm KK } =
3$\,TeV) for {\bf 35} of $SU(4)$, and none for {\bf 10} of $SU(4)$. Note that the direct detection bound for
$M_{\rm KK } = 3$\,TeV with {\bf 10} of $SU(4)$ is not shown because for this case the entire range of DM mass
considered here is ruled out by the central value of the bound. For $M_{\rm KK } = 4$\,TeV with
{\bf 35} of $SU(4)$ the bound is not shown because in that case the central value (DM mass$\sim 40$\,GeV) is below the smallest mass
shown in the plot (i.e., effectively the direct detection bound is weak in this case so that the relic density constraint is
more important).
The reason why the model with DM in {\bf 10} of $SU(4)$ turns out to be more constrained by direct detection than the model with {\bf 35} of $SU(4)$ is due to the fact that $T_{ 3 R } ^{ \nu^{ \prime } }$ in the former model is twice as large as in the latter model, while the $\theta^{ \prime }$ dependence of the effective DM coupling to $Z$ (which controls the interaction with the nuclei) cancels
(see table~\ref{couplingscust1}).
We do not include contribution to direct detection from the radion exchange here
since it is highly model-dependent, and can be easily made sub-dominant for suitable choice of radion mass and $\Lambda_r$.

However, as mentioned already, DM coupling to $Z^{\prime}$ vanishes for the simplest unified model of  ${\bf
15}$ of $SU(4)$, which means that $t$-channel exchange of $Z$ also becomes irrelevant for direct detection
bounds. Hence, for this model, $t$-channel exchange of the radion is the single most important channel, and
direct detection bounds can give information for $\Lambda_r$ and radion mass.
In the CM frame, in the non-relativistic limit, the elastic cross-section for radion exchange is approximately
given as
\beq\label{eq:DD} \sigma\left(\nu' N \to \nu' N \right) \approx \frac{M_{\nu'}^2 \lambda_N^2}{4\pi v_{\rm rel}
\Lambda_r^2} \frac{(|{\bf p}_{\nu'}|^2 + m_N^2)}{(t-m_r^2)^2} \,, \eeq
where $|{\bf p}_\nu'| \approx M_{\nu'} v_{\nu'}$, $v_{\nu'}\sim10^{-3}$ is the DM velocity in the CM frame,
$m_N \approx 1$~GeV is the nucleon mass, $\lambda_N/\sqrt{2}$ is the effective $r\bar N N$ coupling, and $t$ is
the Mandelstam variable, which can be ignored compared with $m_r^2$ in the radion propagator.
For the radion-nucleon coupling we find the typical magnitude $\lambda_N\sim 10^{-6}$ which includes the radion
tree-level coupling to light quarks ($u$, $d$, $s$) and gluon, and the heavy-quark-loop two-gluon couplings,
with the leading parametric dependence $\lambda_N\propto m_N/\Lambda_r$\,. A sub leading dependence on the mass
of radion arises because the radion couplings to gauge boson pairs depend on $m_r$.
All in all, the model parameters enter the direct detection computation in the following way
\beq\label{eq:DDp}\sigma\left(\nu' N \to \nu' N \right)\propto\frac{M_{\nu'}^2m_N^4}{\Lambda_r^4m_r^4}\,.\eeq
For heavy DM, one factor of $m_N^2$ should be replaced by a factor of $M_{\nu'}^2$ arising from the large momentum carried by the heavy $\nu_R'$ and entering the nominator of (\ref{eq:DD}). Finally, note that while Eq.~(\ref{eq:DD}) provides a reasonable approximation, useful for obtaining an analytical understanding of the parameter dependencies of the direct detection constraints, in practice we incorporate our model into the micrOMEGAs~\cite{Belanger:2006is,Belanger:2008sj} package and compute the direct detection bound numerically. We find that the numerical results follow the parametric dependence given in Eq.~(\ref{eq:DDp}) rather well.

Our results are illustrated in Fig.~\ref{fig:DD_SE_Mdm_Mr_KK3}, where the direct detection constraints are superimposed on top of the SE factor. Taking the direct detection limits on face value, we find that a very light radion of $m_r\lsim20$ GeV is already excluded by both the CDMS and Xenon experiments. The CDMS limit disfavor $m_r$ of up to $\sim40$ GeV for DM mass as high as
$3$\,TeV. The entire region of the remaining parameter space, where our analysis is valid, will be probed by upcoming experiments, such as Super-CDMS and Xenon 1-ton~\cite{future}.


\subsection{Indirect detection, simplest fully
unifiable model}\label{ssec:CRprod}

In the following sections we evaluate the implications of our framework to various CR species.
As we have discussed above, models where the DM is not custodially protected are in tension
with direct detection experiments or lead to too low relic density.
We therefore focus on the unifiable model where the DM$-Z'$ coupling vanishes.
To facilitate the discussion we introduce two benchmark models and
study the resulting CR injection spectra and rates. We then move on to signatures in photons and neutrinos. High energy photon and neutrino observations constrain the DM annihilation cross section, weighted by the integral of the DM number density--squared along the line of sight of the experiment.

Proceeding to antiprotons, we note that the astrophysical background is constrained by existing CR data. Subject to few general assumptions, the effect of propagation in the Galaxy can be accounted for at the cost of introducing a single additional fuzz factor. The antiproton analysis is, in this sense, as predictive as the analysis of photon signals where the analogous fuzz factor is contained in the line of sight integral.
Lastly, we turn to the more involved case of positron signals and briefly discuss the
injection rate of $e^+/\bar p$ for which the astrophysical background is somewhat easier to interpret.
\subsubsection{Benchmark models and CR injection spectrum}\label{sssec:inj}
%

Following the discussion of the relic density, Sommerfeld enhancement and direct detection bounds, we focus
here on two viable benchmark model points characterized by different values of DM and radion masses
which result in turn with different annihilation spectra. We keep fixed the value of the Z' mass, $m_{Z'}=3$ TeV.
The benchmark models are defined as follows

%
\begin{description}
  \item[Model L:] $M=600 \ {\rm  GeV}, \ m_{r} > 40 \ {\rm GeV}$, which corresponds to the LH circle
  on  Fig. \ref{fig:relic_density_Mdm_GUT}. In principle one can obtained a sizable SE while decreasing $\Lambda_R$, however, in this case we find tension with
  direct detection bounds (from radion exchange).

  \item[Model H:] $M= 2400 \ {\rm GeV}, \ m_{r}=\cal{O}$(100) GeV, which corresponds to the RH circle
  on  Fig. \ref{fig:relic_density_Mdm_GUT}. In this case theres a wide range of radion masses and corresponding $\Lambda_R$
which yield a sizable SE and consistent with direct experiment.
  \end{description}
  In both cases the annihilation is dominated by $\bar{\nu'}\nu'\to t_R \bar t_R$ via t-channel $X_s$ exchange,
  the couplings are given in table~\ref{couplingscustGUT} and according to Eq.~(\ref{MDMc}) which
  link the top compositeness with the DM mass.
%

\begin{figure}[hbp]
%
%
\includegraphics[width=\linewidth]{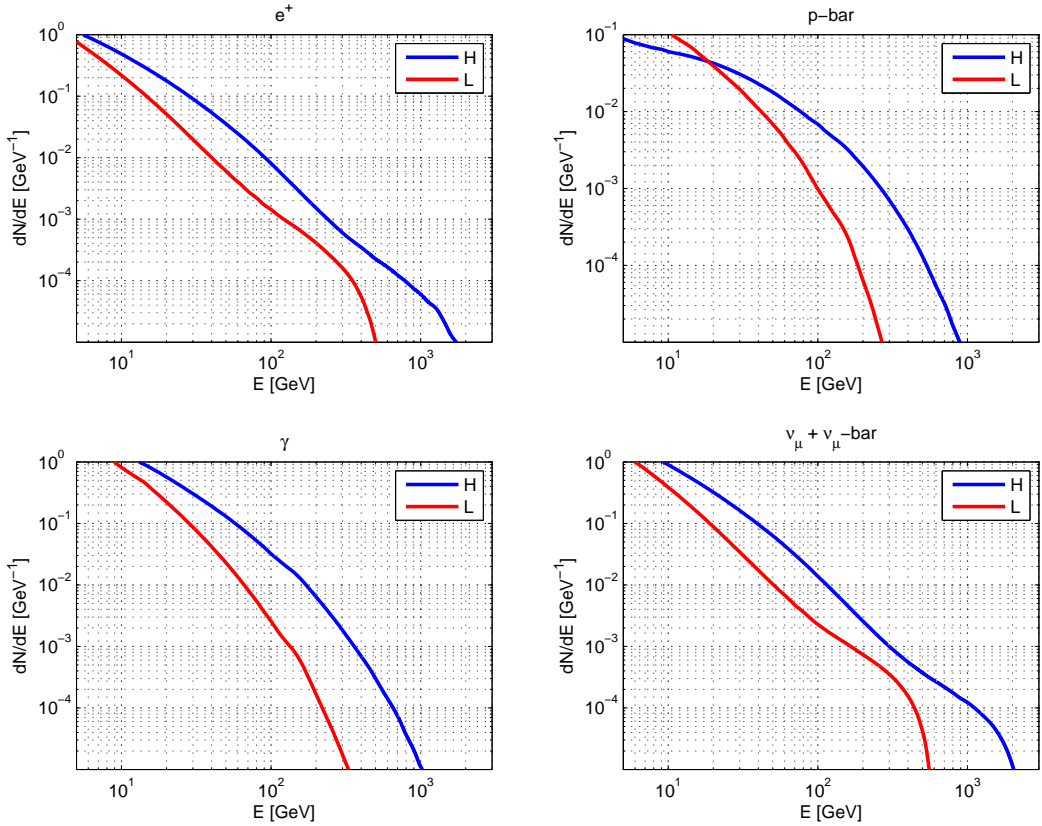}
\caption{Decayed final state annihilation spectra for the two benchmark models.} \label{fig:ModelsSpec}
\end{figure}

The CR injection spectra of stable final states are plotted in Figure~\ref{fig:ModelsSpec} for the various
benchmark points. These spectra, together with the DM mass and Sommerfeld enhancement factor serve as the particle physics input required for the calculation of indirect detection signals.
%

\subsubsection{CR production rate}\label{sssec:prod}

The production rate of a cosmic ray specie $\alpha$ due to annihilation of Dirac fermion DM at a given spatial
position $\vec r$ in the Galaxy is given by
\beq\label{eq:Q}Q_{_{\alpha,\rm DM}}(E,\vec r)=\frac{1}{4}n^2(\vec r)\frac{d\sigma v(DMDM\to\alpha)}{dE}\,.\eeq
Here $n(\vec r)$ is the total DM number density (particle+antiparticle) and $\frac{d\sigma
v(DMDM\to\alpha)}{dE}$ is the differential velocity-weighted annihilation cross section for the production of the specie
$\alpha$. It is convenient to work with dimensionless quantities,
\beq\label{eq:var}\begin{split}&\epsilon=\frac{E}{\rm GeV}\,, \ \ \ M_1=\frac{M}{\rm TeV}\,, \ \ \ \overline{\sigma
v}=\frac{\sigma v}{6\cdot10^{-26} \, {\rm cm}^3{\rm\, s}^{-1}}\,, \ \ \ n_o(\vec r)=\frac{n(\vec r)}{n(\vec
r_{\rm sol})}\,,
\end{split}\eeq
where $M$ is the DM mass, $\sigma v$ is the total velocity-weighted annihilation cross
section, $\vec r_{\rm sol}\approx8.5$\,kpc is the distance between the solar system and the Galactic center and $n(\vec r_{\rm sol})=0.3{\rm\,cm}^{-3} \text{GeV}/M$ is the DM number density in the local halo. For the local halo mass density, we adopt
a value of $\rho_{DM}(\vec r_{\rm sol})=0.3 \, {\rm GeV}\,{\rm cm}^{-3}$. Order one deviations for this number are possible both on average and due to local clumps, and go through to the computed CR flux. With the definitions (\ref{eq:var}), the CR production rate can be written as
\beq\label{eq:QDM}Q_{_{\alpha,\rm DM}}(\epsilon,\vec r)&=Q_{_{\alpha,\rm DM}}(\epsilon,\vec r_{\rm sol})\times n_o^2(\vec r)\,,\eeq
with the local injection rate
\beq\label{eq:localQDM}
Q_{_{\alpha,\rm DM}}(\epsilon,\vec r_{\rm sol})&=1.3\cdot10^{-33} \ \frac{\overline{\sigma
v}}{M_1^2} \ \frac{dN_\alpha}{d\epsilon} \ \ \ {\rm cm}^{-3}{\rm \,s}^{-1}{\rm\,GeV}^{-1}\,,\eeq
and where $\frac{dN_\alpha}{d\epsilon}$ is the differential number of stable final state particles of specie
$\alpha$ emitted per annihilation event. In writing Eq.~(\ref{eq:QDM}) we have neglected the spatial dependence in the Sommerfeld enhancement~\cite{Robertson:2009bh}. As, in this paper, we do not attempt to provide a detailed description of the spatial features of the DM annihilation signal, we neglect this possible complication throughout the discussion.

The rate of DM annihilation is proportional to the number density squared, and so the results, in particular as
concerns photon and neutrino flux from the Galactic center region, depend on the assumed profile.
The latest N-body simulations, including only DM and no baryons, point to DM halo profiles with a cusped
central region. However, the inner zone of a few hundreds of parsecs from the center remains uncertain. In
addition, the effect of baryons may be significant at the central region and its impact on the DM distribution
is far from understood. Baryons were argued to either increase the inner cusp, or actually smooth it out,
resulting with a cored profile~\cite{RomanoDiaz:2008wz,Sellwood:2008bd}. In this work we analyze both cusped
and cored DM halo profiles. The examples we consider are the cusped NFW~\cite{Navarro:1995iw}
and the cored isothermal sphere~\cite{Bahcall80} (denoted below by ISO). We do not attach special significance to any particular
profile but rather lay out the consequences of each case regarding indirect detection prospects for our
framework. The radial dependence of the halo distributions is,
\beqa\label{eq:DMhalo}\begin{split}
\text{NFW:} \ & \ \frac{\rho(r)}{\rho(r_{\rm sol})}&=&\frac{r_{\rm sol}}{r}\left(\frac{1+r_{\rm sol}/r_s}{1+r/r_s}\right)^2, \ \ r_s=20 \, \text{kpc}\,,\\
%
%
\text{ISO:} \ & \ \frac{\rho(r)}{\rho(r_{\rm sol})}&=&\frac{1+r_{\rm sol}^2/r_s^2}{1+r^2/r_s^2}, \ r_s=5 \,\text{kpc}\,.
\end{split}\eeqa
%

\subsubsection{Photons and neutrinos}\label{sssec:phot}

The incoming flux of photons or neutrinos per unit solid angle is obtained by integrating the production rate
along the line of sight in a given direction $\Omega$ in the sky,
\beqa\label{eq:j}\begin{split} j(\Omega,\epsilon)d\Omega&=d\Omega\int_{\mathrm{l.o.s.}} drr^2
\frac{Q_{\gamma(\nu),DM}(\epsilon,\vec r)}{4\pi r^2}\\
&=Q_{\gamma(\nu),DM}(\epsilon,\vec r_{\rm sol})\times\frac{d\Omega}{4\pi}\int_{\mathrm{l.o.s.}}
drn_o^2(\vec r)\,.
\end{split}\eeqa
Gamma ray observatories report limits on $\bar j(\Omega,\epsilon)$, which is defined by averaging (\ref{eq:j})
over the acceptance $\Delta\Omega$ of the experiment,
\beq\label{eq:jbar}\bar j(\Omega,\epsilon)=\frac{1}{\Delta\Omega}\int_{\Delta\Omega}d\Omega
j(\Omega,\epsilon)\,.\eeq
The observed photon flux depends on the local injection rate, up to a single over all model dependent factor
given by the line of sight integral, which encodes the DM distribution.

We derive gamma ray--based model constraints from the following data sets, provided by the HESS imaging air
Cherenkov detector:
\begin{itemize}
  \item {HESS observations of the Galactic Center (GC)~\cite{Aharonian:2006wh}:\\
The GC data set corresponds to the inner $0.1^\circ$ of the GC gamma-ray source, HESS J1745-290. The energy
range was $E_\gamma>160$\,GeV. Considering the uncertainties involved in the calculation, we find it sufficient
for our purpose to use the power law fit reported by the HESS collaboration, $\bar j \propto E^{-\Gamma}$ with
$\Gamma=2.25\pm0.04(\mathrm{stat})\pm0.10(\mathrm{syst})$. The normalization is defined from the reported value
of the integrated flux above 1 TeV, $\int_{_{_{1 \mathrm{TeV}}}}dE \bar j
\Delta\Omega=\left[1.87\pm0.10(\mathrm{stat})\pm0.30(\mathrm{syst})\right]\cdot10^{-12} \, {\rm cm}^{-2}{\rm \,s}^{-1}\,$. For
definiteness, we use the central values for the reported flux and impose that the photon flux resulting from DM
annihilation does not exceed it in order to derive constraints on the model.}

  \item {HESS observations of the Galactic Ridge area (GR)~\cite{Aharonian:2006au}:\\
The GR data set corresponds to an observation of the rectangular angular patch $|l|<0.8^\circ, \
|b|<0.3^\circ$, from which the spectral components of the sources HESS J1745-290 and G0.9+0.1 \footnote{see
\cite{Aharonian:2005br} for the details of the source G0.9+0.1} were subtracted. The energy range was
$E_\gamma>170$\,GeV. We use the power law fit reported by the HESS collaboration, $\bar
j=k\left(\frac{E}{\rm TeV}\right)^{-\Gamma}$ with $\Gamma=2.29\pm0.07(\mathrm{stat})\pm0.20(\mathrm{syst})$ and the
normalization $k=\left[1.73\pm0.13(\mathrm{stat})\pm0.35(\mathrm{syst})\right]\cdot10^{-8} \
{\rm \,TeV}^{-1}{\rm \,cm}^{-2}s^{-1}{\rm\,sr}^{-1}$. We use the central values for the reported flux and impose that the photon flux
resulting from DM annihilation does not exceed it in order to derive constraints on the model.
 }

\end{itemize}
Data from the Fermi-LAT satellite--borne detector has recently become available. We analyzed the preliminary results presented in~\cite{FermiTeVPA} for the Galactic center region. This data constrains the lower energy part of the spectrum and, for model L with a cuspy DM profile, is competitive with the HESS data.

\begin{figure}[hbp]\begin{center}
\hspace{-2cm}
\includegraphics[width=150mm,height=150mm]{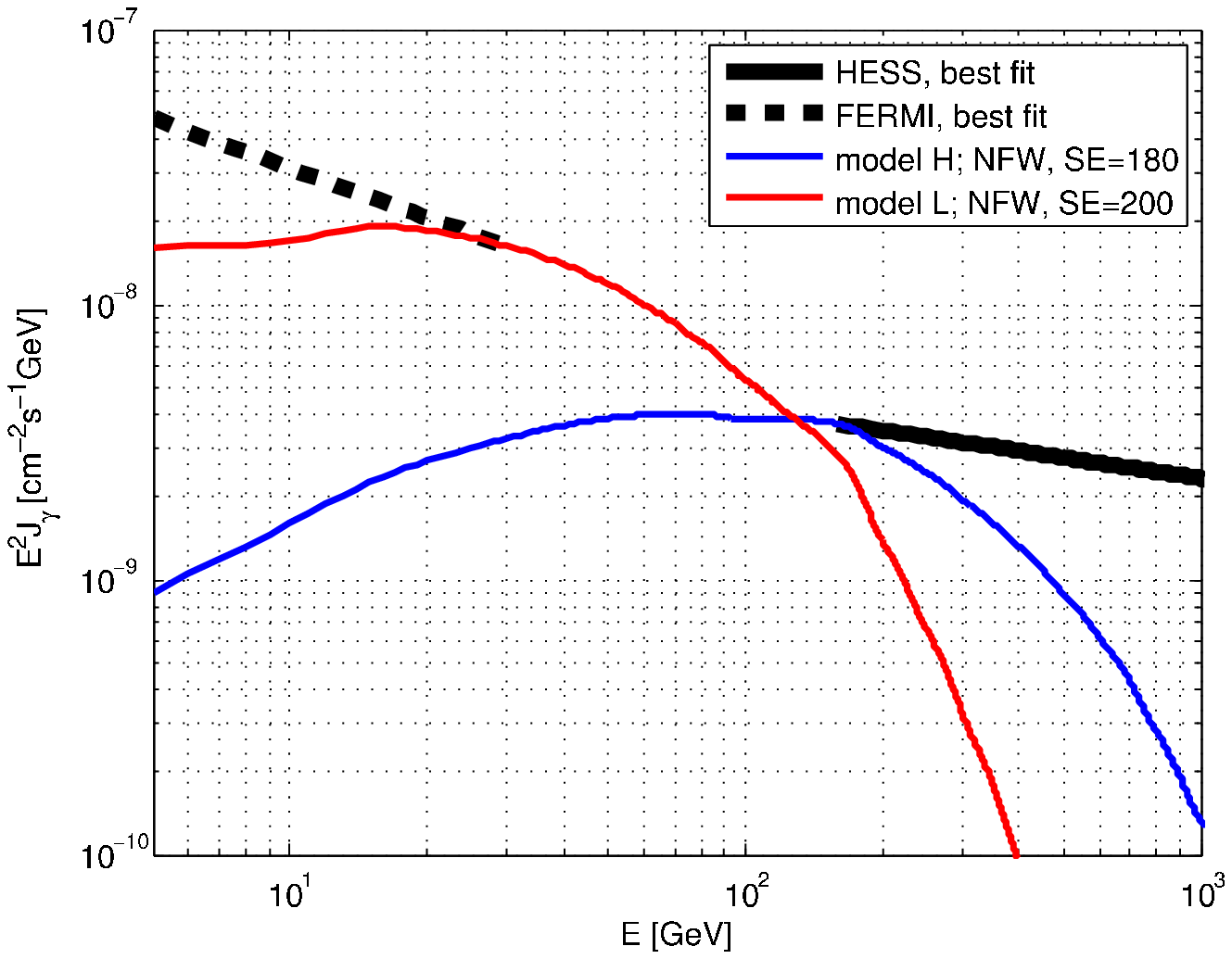}
\caption{Gamma ray constraints from FERMI and HESS.}\label{fig:hessfermi}
\end{center}
\end{figure}

The situation is illustrated in Fig.~\ref{fig:hessfermi}, in which we plot the GC data set of FERMI and HESS vs
model signals, evaluated with an NFW DM halo profile and the maximal Sommerfeld factor allowed by the GC data
set. (note that, for model H, the HESS GR data set is in fact more constraining and a value of $SE=180$, used
in the figure for illustration, is excluded.)

Limits on the neutrino flux arise from measurements of the neutrino--induced muon flux in neutrino detectors.
For DM mass M, the flux of muons at the detector is given by
\beq\label{eq:muF}\phi_\mu= \int_{\epsilon_{th}}^Md\epsilon_\mu\int_{\epsilon_\mu}^Md\epsilon_\nu\bar n_N\bar
j_{\nu_\mu}(\Omega,\epsilon)\Delta\Omega\left[\frac{d\sigma^{\nu N}}{d\epsilon_\mu}+\frac{d\sigma^{\bar\nu
N}}{d\epsilon_\mu}\right]L(\epsilon_\mu,\epsilon_th)\,.\eeq
Here $j_{\nu_\mu}(\Omega,\epsilon)$ is the muon-neutrino flux at the earth, which equals the anti-neutrino flux
in the case of DM annihilation and is obtained from Eq.~(\ref{eq:jbar}) (We use Tri-Bimaximal neutrino mixing
for definiteness). The differential cross sections are given by
\beq\label{eq:dsignumu} \frac{d\sigma^{\nu N}}{d\epsilon_\mu}=\frac{2G_F^2\bar
m_N}{\pi}\left[a_1+a_2\left(\frac{\epsilon_\mu}{\epsilon_\nu}\right)^2\right]\,,\eeq
with $a_1\approx0.2,a_2\approx0.05$ for neutrino--nucleon CC scattering and the same with $a_{1,2}$
interchanged for the antineutrino--nucleon case. The muon range in the rock beneath the detector is
\beq\label{eq:Lmu}L(\epsilon_\mu,\epsilon_{th})=\frac{1}{\rho\beta_\mu}\ln
\left(\frac{\alpha_\mu+\beta_\mu\epsilon_\mu}{\alpha_\mu+\beta_\mu\epsilon_{th}}\right)\,,\eeq
where $\epsilon_{th}=1.6$\,GeV is the threshold energy for detection, and $\alpha_\mu\approx2\cdot10^{-3}\,
{\rm GeV\,}{\rm cm}^2{\rm \,g}^{-1}$ and $\beta_\mu\approx3\cdot10^{-6}{\rm \,cm}^2{\rm \,g}^{-1}$ are the muon energy loss coefficients. For the
target material we consider a nucleon mass $\bar m_N=m_p$, and the nucleon number density is given by $\bar
n_N=\rho/\bar m_N$.

We derive neutrino--based model constraints from the upper limits on the upward through-going muon flux,
measured at Super Kamiokande (SK)~\cite{Desai:2004pq}. We use the 95\%CL limits quoted in~\cite{Mardon:2009rc},
where the line of sight integrals were also given for angular acceptances of $3^o-30^o$ and various DM halo
profiles.

In Table~\ref{tab:Bmxphotnu} we summarize the photon and neutrino constraints.
Regarding observations of the Galactic center region, the line of sight integral
depends on the assumed DM halo profile as well as the angular resolution
of the experiment. Small changes in the halo profile around the poorly-known
central regions of the Galaxy result with significant variations in 
the predicted flux~\cite{Meade:2009rb}. For the cored profile, large cancellations can occur due to background
subtraction, and the resulting bound becomes weak~\cite{Mardon:2009rc} in comparison with antiproton and
neutrino constraints. In this case, for the HESS analysis we report the bound without accounting for these
cancellations (given inside square brackets in Table~\ref{tab:Bmxphotnu}), such that the optimal performance can
be assessed.

Coming back to the Fermi data, we note that a part of the power of these measurements lies in the complete coverage of the sky. As a result, strong constraints can be derived also for cored DM profiles, which were effectively unconstrained by earlier measurements. A more complete treatment of the new Fermi data, which included the same final state annihilation products as in our model, was very recently provided in~\cite{Papucci:2009gd,Cirelli:2009dv}. The analysis of these references is in good agreement with ours for observations of the Galactic center, but as expected, it presents much stronger bounds for the cored profile. In particular,
according to~\cite{Papucci:2009gd,Cirelli:2009dv}, the SE for our model L(H) cannot exceed $\sim500(10^3)$ in the ISO profile scenario. For an NFW profile, the SE for model L(H) is limited below $\sim50(100)$.

Putting all together and including the results of~\cite{Papucci:2009gd,Cirelli:2009dv}, we find that the neutrino bounds are sub-dominant in comparison with the new photon data, for any DM profile. Finally, note that with a realistic treatment of the backgrounds, the bounds we
apply are likely to tighten by a factor of at least a few, implying that the SE factor for our models will probably be
limited to a few tens (hundreds) in case a cusped (cored) DM profile is adopted. As we show below, such value of the SE is still sufficient to produce interesting antiproton signatures.

\begin{table}[t]
\caption{Upper bounds on the Sommerfeld enhancement factor SE, resulting from HESS and FERMI $\gamma$ and SK
$\nu$ constraints. Square brackets refer to optimal
background subtraction with the HESS resolution. For $\nu$ we report the result corresponding with the most
constraining opening angle for the SK analysis. In case the bounds are weaker than $10^4$, we keep only the
order of magnitude. We also quote the analyses of Fermi constraints, provided in~\cite{Papucci:2009gd,Cirelli:2009dv}; see text for details.} \label{tab:Bmxphotnu}
\begin{center}
\begin{small}
\begin{tabular}{|l|c|c|c|c|c|} \hline\hline
\rule{0pt}{1.2em}%
 & GC, HESS $\gamma$  \ \cite{Aharonian:2006wh} &  GR, HESS $\gamma$  \ \cite{Aharonian:2006au} &
 GC, FERMI $\gamma$  \ \cite{FermiTeVPA} & FERMI $\gamma$\, \cite{Papucci:2009gd,Cirelli:2009dv} & SK $\nu$ \ \cite{Desai:2004pq}\cr \hline\hline
 & NFW\, | ISO & NFW\, | ISO & NFW\, | ISO & NFW\, | ISO & NFW\, | ISO
\cr\hline
L & $260$\, | $[10^5]$ & $310$\, | $[10^4]$ & $200$\, | $10^7$ & $50$\, | $500$ & $4\cdot10^3$\, | $10^4$
\cr\hline
H & $180$\, | $[10^5]$ & $130$\, | $[10^4]$ & $900$\, | $10^8$ & $100$\, | $10^3$ & $5\cdot10^3$\, | $10^4$
\cr\hline
\end{tabular}
\end{small}
\end{center}
\end{table}

\subsubsection{Antiprotons}\label{sssec:pbar}

The PAMELA experiment has recently measured the antiproton to proton flux ratio~\cite{Boezio:2008mp}. The
reported antiproton fraction does not show deviations from the expected result, based on secondary production
by pp and spallation interactions of primary CRs with inter-stellar medium (ISM). Nevertheless, DM annihilation can contribute a
primary component to the CR antiproton flux, with a production rate density given by Eq.~(\ref{eq:QDM}). This
contribution must be small at the currently explored energies, but could in principle reveal a peak at
$\gsim100$\,GeV energies, soon to be measured by the PAMELA and (hopefully) AMS02 \cite{Alpat:2003pn} experiments.

Cosmic ray antiprotons have long been considered as a good channel to detect exotic sources (see {\it
e.g.}~\cite{pbarDM}). At high energies (E\,$>10$\,GeV) the background can be determined from the CR nuclei
data~\cite{Gaisser:1991vn,Simon:1998,Katz:2009yd}, leaving significant predictive power. As concerns the DM
contribution, analyses in the literature were based on detailed propagation models. Such propagation models
typically include additional free parameters which reduce the predictive power of the analysis. Here we show
that the antiproton flux can be computed in a model independent manner, at the cost of introducing one free
parameter to the calculation. This parameter is an energy independent effective volume factor, encoding the
different spatial extensions of the DM and the spallation sources. The fact that only one free parameter is
introduced makes the antiproton channel as predictive as the photon and neutrino channels, for which
propagation in the Galaxy is trivial.

The approach we adopt is based on the fact that high energy antiprotons above a few GeV suffer only small
energy losses as they travel through the Galaxy, and on the fact that the secondary antiproton flux up to
E\,$\sim300$\,GeV can be computed in a model independent manner, based on the existing CR nuclei data~\cite{Gaisser:1991vn,Simon:1998,Katz:2009yd}. To proceed, we need the following ingredients.

The first ingredient concerns the propagation in the Galaxy. Define the quantity
$G(\epsilon,\epsilon_S;\vec r,\vec r_S)$\,, encoding the propagation of CR antiprotons in the Galaxy as follows
\beq\label{eq:pbsteadystate} n_{\bar p}(\epsilon,\vec r)=\int d^3r_S\int d\epsilon_SQ_{\bar p}(\epsilon_S,\vec
r_S)G(\epsilon,\epsilon_S;\vec r,\vec r_S)\,,\eeq
where $Q_{\bar p}(\epsilon_S,\vec r_S)$ is the injection rate density at energy $\epsilon_S$ at the point $\vec
r_S$\,, the spatial integral contains the confinement volume of Galactic CRs and $n_{\bar p}(\epsilon,\vec r)$ is
the antiproton density at some point $\vec r$ in the Galaxy~\footnote{For simplicity we did not introduce a time label in Eq.~(\ref{eq:pbsteadystate}), which could account for deviations from steady state. Provided that the explicit assumptions we make hold, adding time dependence to the problem would not change our basic result.}. The negligible energy change of antiprotons above
10 GeV implies that $G(\epsilon,\epsilon_S;\vec r,\vec r_S)\propto\delta(\epsilon-\epsilon_S)$\,. We now make the assumption that $G$ is separable, {\it i.e.} that
\beq G(\epsilon,\epsilon_S;\vec r,\vec r_S)=\delta(\epsilon-\epsilon_S)g(\epsilon)\bar G(\vec r,\vec r_S)\,.\eeq

The second ingredient concerns the secondary source spectrum. We assume that the injection spectrum (not rate)
of secondary antiprotons has a homogeneous distribution in the Galaxy. In practice this assumption amounts to
demanding that spatial variations in the spectrum of primary CRs are small, at least in the regions from which
most of the secondary antiprotons observed locally are generated. Under this assumption the secondary source
term is separable,
    \beq Q_{\bar p,\rm sec}(\epsilon,\vec r)=Q_{\bar p,\rm sec}(\epsilon,\vec r_{\rm sol})\times q_{\rm sec}(\vec r)\,,\eeq
where $Q_{\bar p,\rm sec}(\epsilon,\vec r_{\rm sol})$ is the local secondary injection rate.

Under the above assumptions, the local density ratio between the primary
and secondary components takes the form:
\beq\frac{n_{\bar p,\rm DM}(\epsilon,\vec r_{\rm sol})}{n_{\bar p,\rm sec}(\epsilon,\vec r_{\rm sol})}=f_V\frac{Q_{_{\bar
p,\rm DM}}(\epsilon,\vec r_{\rm sol})}{Q_{\bar p,\rm sec}(\epsilon,\vec r_{\rm sol})},\eeq
with the energy independent volume factor, \beq f_V=\frac{\int d^3rq_{DM}(\vec r)\bar G(\vec r_{\rm sol},\vec
r)}{\int d^3rq_{\rm sec}(\vec r)\bar G(\vec r_{\rm sol},\vec r)}\,.\eeq
For a DM annihilation source, we have $q_{DM}(\vec r)=n_o^2(\vec r)$\,. We can write the antiproton to proton
flux ratio as follows,
\beqa\label{eq:BFnn}\begin{split} \frac{J_{\bar p}(\epsilon,\vec r_{\rm sol})}{J_p(\epsilon,\vec r_{\rm sol})}
&=\left(\frac{J_{\bar p}(\epsilon,\vec r_{\rm sol})}{J_p(\epsilon,\vec r_{\rm sol})}\right)_{\rm
sec}\times\left[1+f_V\frac{Q_{_{\bar p,\rm DM}}(\epsilon,\vec r_{\rm sol})}{Q_{_{\bar p,\rm sec}}(\epsilon,\vec
r_{\rm sol})}\right]\,.\end{split}\eeqa

The first factor on the right hand side is the secondary antiproton to the primary proton flux ratio. This
quantity is constrained by the B/C data, leaving no free parameters. We conclude that under some general assumptions, the antiproton to proton flux ratio including a DM contribution can be computed
based on the relatively well constrained local injection rates and only one additional parameter, $f_V$,
encapsulating all the details of the propagation. A naive estimate suggests $f_V\sim L/h\sim10-100$\,, where $L\sim1-10$\, kpc is the assumed half width of the CR propagation volume and $h\sim100$\, pc is the half width of the Galaxy gaseous disc.

The class of models for which Eq.~(\ref{eq:BFnn}) holds includes the disc+halo diffusion
model with a homogeneous diffusion coefficient~\cite{Ginzburg:1990sk, Strong:2007nh}\footnote{Of course, the class of models for which Eq.~(\ref{eq:BFnn}) holds, includes also the well known leaky box model~\cite{Schlickeiser:1985}.}. In
Appendix~\ref{app:pbar} we use this model as a concrete example, deriving the precise realization of
Eq.~(\ref{eq:BFnn}). We find, as expected, $f_V$ in the range $\sim10-100$, depending mainly on the size of the
CR confinement halo with an order one correction depending on the DM distribution.

\begin{figure}[hbp]\begin{center}
\hspace{-2cm}
\includegraphics[width=150mm,height=150mm]{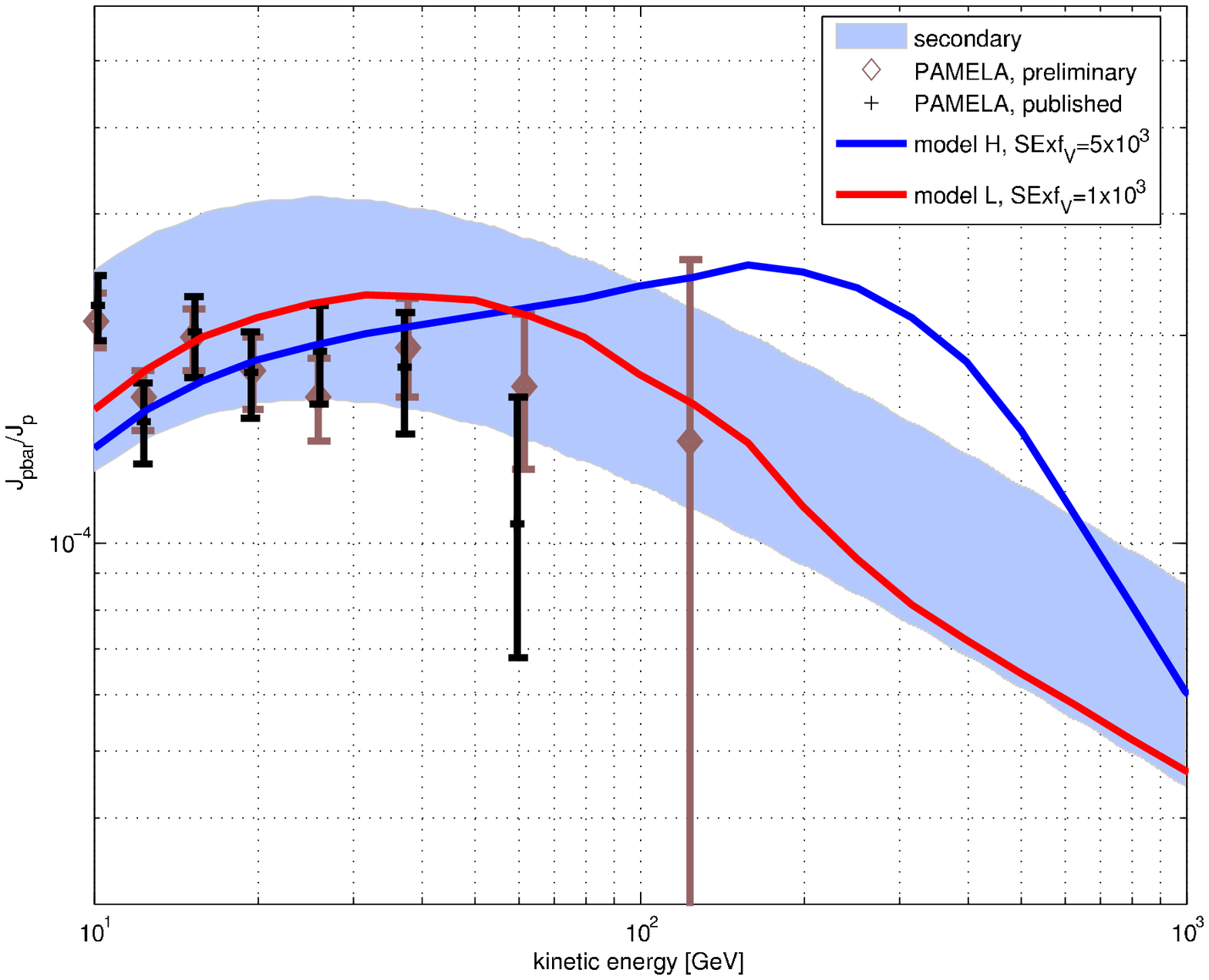}
\caption{The antiproton to proton flux ratio with a DM component, corresponding to our benchmark models. The
curves including a DM contribution are obtained by suppressing the background prediction to $75\%$ of its
central value, and boosting the local DM injection rate by the factor ${\rm SE}\times f_V$. This factor encodes
the combination of propagation, via the volume factor $f_V$, and of the Sommerfeld enhancement SE. The shaded
region indicates an $40\%$ uncertainty estimate for the background calculation. Data points are from published
and preliminary PAMELA data.\label{fig:BFLIB}}
\end{center}
\end{figure}

In Fig.~\ref{fig:BFLIB} we plot the antiproton to proton flux ratio with a DM component, corresponding to our
benchmark models. The curves including DM contribution are obtained by suppressing the pure background term to
$75\%$ of its central value (we find that a similar suppression also best describes the data with only the
background component), and boosting the DM term by the factor SE\,$\times f_V$, indicated in the plot. The
shaded region denotes an $40\%$ uncertainty estimate for the background calculation \cite{Simon:1998}. Data
points are taken from published~\cite{Boezio:2008mp} and preliminary~\cite{PAMprelim} PAMELA data. As
illustrated in Fig.~\ref{fig:BFLIB}, a future $\bar p$ signal can arise for $m_{\rm DM}\gsim 1\,$TeV, with
SE\,$\times f_V\gsim10^3$. As a volume factor $f_V>10$ is envisioned, the requirement on the Sommerfeld factor
is SE\,$\gsim100$, easy to obtain in our model with a 100 GeV radion. The TeV scale for the DM mass, roughly
above which the resulting antiproton feature can be pushed higher than existing constraints to provide a future
signal, is indicated by the vertical golden line of Fig.~\ref{fig:relic_density_Mdm_GUT}.

Concerning the astrophysical background calculation depicted in Fig.~\ref{fig:BFLIB}, a comment is in order.
Extending the background prediction all the way to E\,$\sim$\,TeV requires extrapolation of the CR grammage
(provided in Appendix~\ref{app:pbar}, Eq.~(\ref{eq:X})) beyond the 200-300 GeV, where reliable data exists
\cite{Engelmann:1990zz}. While there are indications that the grammage used in Fig.~\ref{fig:BFLIB} persists to
TeV energies~\cite{Ahn:2008my}, the issue is not currently settled~\cite{Zatsepin:2009zr}. We anticipate that
with improved compositional CR data, extending to TeV energies, an updated model independent prediction for the
$\bar p/p$ ratio will become directly available along the same lines described
above~\cite{Gaisser:1991vn,Simon:1998,Katz:2009yd}. Eq.~(\ref{eq:BFnn}) will then become useful up to TeV
energies.

%

\subsubsection{Electrons, positrons and the positron to antiproton flux ratio}\label{sssec:pos}

Recently the PAMELA collaboration have reported a rise in the positron to electron plus positron
fraction~\cite{Adriani:2008zr}, beginning at E\,$\sim10$\,GeV. The reported rise have induced numerous
publications, suggesting an explanation in terms of DM annihilations or decay. However, before examining exotic
contributions it is necessary to first understand the astrophysical background, which is harder to constrain
than in the antiproton example.

In fact, since secondary positrons are produced by pp and spallation interactions, just like antiprotons, an
upper bound to the positron flux can be obtained model independently, based on the measured CR
grammage~\cite{Katz:2009yd}. Contrasted with the data, this calculation reveals that the rising positron
fraction is not accompanied by any actual positron excess with respect to the model independent upper bound.
One is forced to the conclusion that the rising positron fraction most likely corresponds to an unexpected
spectral behavior of the suppression due to propagation energy losses, denoted here by $f_s$. Using the total
$e^++e^-$ measurements~\cite{Alcaraz:2000bf,DuVernois:2001bb,:2008zzr,Abdo:2009zk} in conjunction with the
PAMELA data, one finds $f_s\approx0.3$ at $E\approx10$ GeV, rising to $f_s\approx1$ at the highest data bin
$E\approx80$ GeV. At E\,$\lsim40$\,GeV, the suppression of the positron fraction can also be compared with the
measured suppression due to decay of the flux of radioactive unstable CR isotopes, such as $^{10}{\rm Be},
^{26}{\rm Al}$ and $^{36}{\rm Cl}$~\cite{Webber:1998,Strong:2007nh}. In particular, measurements of the (purely
secondary) decaying charge to decayed charge ${\rm Be/B}$ extend to rigidity of
$\approx40$\,GV~\cite{Webber:1998}. These measurements suggest a value of $f_s\sim0.3$ for positron energy
$\sim20$\,GeV, in agreement with the actual result and in support of the secondary origin of the detected
positrons.

To summarize, it is our view that the rising positron fraction does not consist an evidence for exotic
components in the positron flux, simply because there does not seem to be any positron excess -- merely an
intriguing suppression pattern. If, however, future data release by the PAMELA mission or other
experiments~\cite{Alpat:2003pn} will establish that the rising behavior persists to $e^+/(e^++e^-)>0.2$ beyond
$\sim$\,100 GeV, a positron excess will indeed be implied, necessitating a primary source.

Both of our benchmark models do not produce a hard lepton flux, hence, no anomaly is predicted in leptonic
channels. This conclusion is supported by the expectation that, due to radiative losses, positrons do not
experience the volume enhancement relevant for antiprotons. However, as argued above, since the background
distribution (both of primary electrons and secondary electrons and positrons) is largely unknown, we proceed
to discuss and analyze, in the following part, a more robust observable related to secondary to secondary flux
ratio.

In terms of theoretical uncertainties, the positron to antiproton flux ratio is a clean discriminator between a
secondary astrophysical production mechanism to any other hypothetical source. The reason for this is that
secondary antiprotons and positrons are produced by the same mechanism, namely $pp$ and spallation interactions
of primary CRs with ISM. The relative amount of positrons and of antiprotons injected at a given energy depends
on the corresponding branching ratios and, to a lesser extent, on the spectrum and composition of the primary
CRs and the ISM. An examination of the dependence of the positron to antiproton ratio on the spectrum of
primaries was carried out in~\cite{Katz:2009yd}, where this dependence was found to be very mild. Since at high
energies energy losses affect only the positrons and act to suppress the observed flux, and since in the
absence of losses high energy positrons and antiprotons would propagate in a similar way, the positron to antiproton
injection rate ratio forms a robust upper bound on the corresponding flux ratio, relatively immune to propagation details.

The spectrum of final state products in DM annihilation may deviate significantly from the corresponding
branching ratios in $pp$ collisions. The existence of a DM component in the CR antimatter flux can
therefore be searched for in the positron to antiproton flux ratio. Finding this ratio above the standard
prediction (based essentially on the branching ratios in $pp$ collisions, with mild compositional corrections)
will provide strong motivation for an exotic contribution.

\begin{figure}[hbp]\begin{center}
\hspace{-2cm}
\includegraphics[width=100mm]{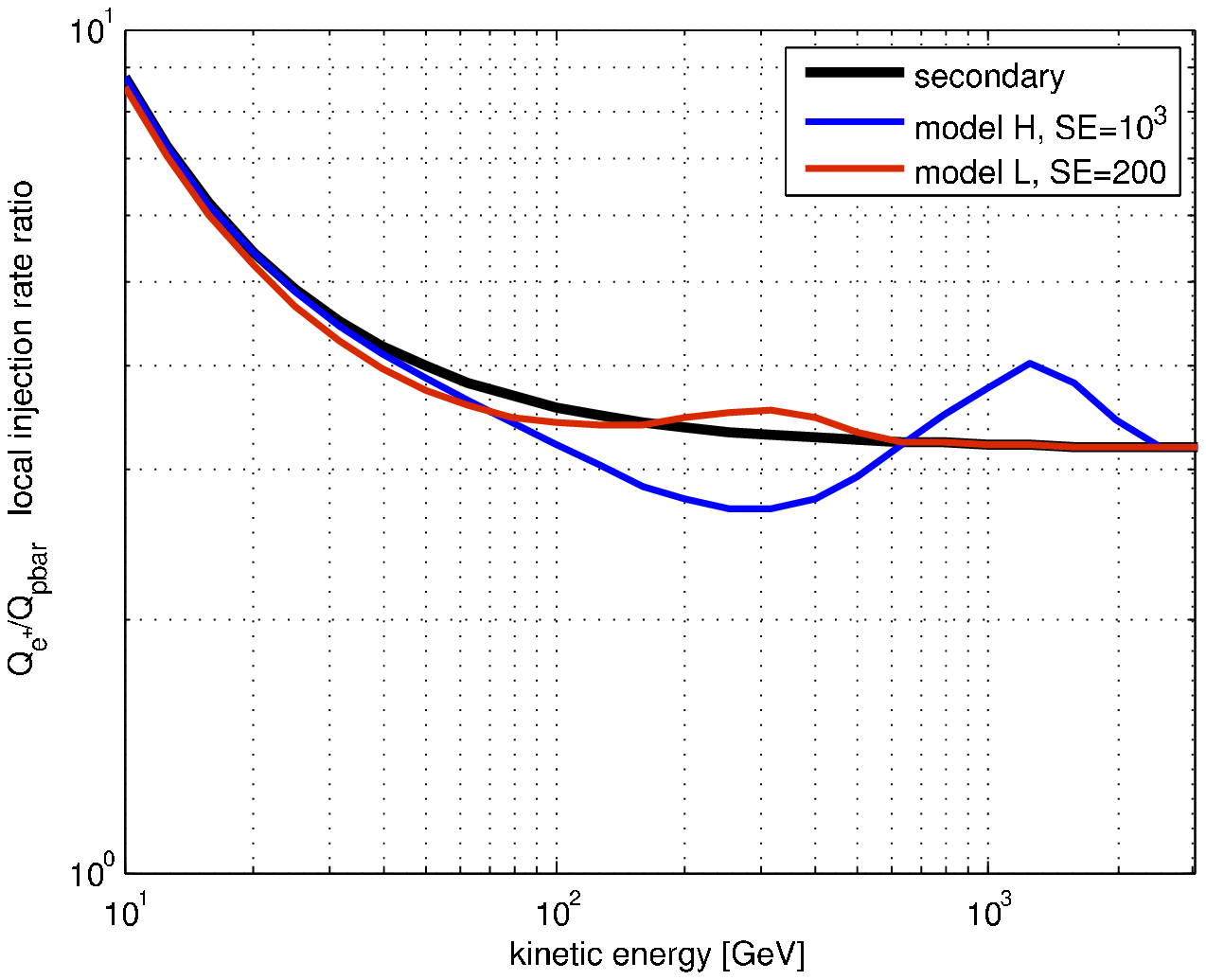}
\caption{Positron to antiproton production rate ratio.}
\label{fig:QposToQpbar}\end{center}
\end{figure}

In figure~\ref{fig:QposToQpbar} we plot the positron to antiproton production rate ratio for our benchmark
models, including the background, compared with the prediction for $pp$ collisions which can be regarded as an
upper bound for the background result. In both models, the ratio lies very close to the astrophysical
background. The conclusion is that it is unlikely, yet not inconceivable, that our models would lead to an
excess in leptonic CR signals.


\section{Radion Collider Phenomenology}\label{sec:coll}
For the region of our model parameter space, where a possible CR signal in
DM annihilation is obtained, a lightish radion with mass in the 100\,GeV range is required.
It implies that the radion may turn out to be the lightest new
particle in our model, likely to be accessible at the LHC.
Various studies on
radion phenomenology have been performed in the past~\cite{Goldberger:1999un,CGRT,GRW,DeWolfe:1999cp,TanakaMontes,CGK,Gunion,Korean, Rizzo}, including recent works where radion dynamics was considered within
realistic models of electroweak breaking with bulk SM fields~\cite{Rizzo,CHL,Toharia}.
For example in the case of radion mass lighter than $2M_{W}$, $r\rightarrow \gamma\gamma$ is a promising channel~\cite{CHL},
which can be also dramatically enhanced in the presence of Higgs-radion mixing~\cite{Toharia}.
For the case of radion mass larger than $2M_{W}$, $WW$, $hh$, $ZZ$,$t \bar{t}$ are the dominant channels which are expected to allow for a discovery at the LHC.
Thus, a discovery of lightish radion at the LHC and future signal at CR experiments, would
yield a support for our class of models.

\section{Conclusions}
\label{conclude}

Indirect signals from DM annihilation in cosmic ray experiments have received a renewed attention. We point out
that models of warped extra dimension can naturally yield a low velocity enhancement of the dark matter (DM)
annihilation via the Sommerfeld effect. The enhancement does not rely on an extra dark sector, but rather is
mediated via an intrinsic component of the theory, namely the radion with a mass at the hundred GeV range. More
specifically, we studied the well motivated framework of a warped grand unified theory (GUT), in which the DM
particle is a GUT partner of the top quark. Based on the Pati-Salam group, we constructed models of partial and
full unification, which accommodate custodial symmetry protection for $Z\to b \bar b$ coupling.
The above construction is consistent with electroweak precision tests for Kaluza-Klein (KK) particles with mass
scale of a few TeV.
In addition, we explore the consequences of a similar custodial symmetry protection of $Z$ couplings to
right-handed (RH) tau's. Such protection enables the RH tau's to be composite, localized near the TeV end of
the extra dimension, hence having a large coupling to KK particles. As an aside, independently of the
requirement for unification, the strong coupling between the KK particles and the composite tau's can lead to
striking LHC signals.

Cosmological and astrophysical aspects of our framework are discussed. We find that the dark matter relic
abundance, as well as direct detection, constrain the viable parameter space of this class of models.
Particularly strong constraints are found in cases where the DM particle couples to the neutral electroweak
sector. Indirect signatures in Galactic cosmic rays (CR) are studied. We focus on robust observables,
relatively immune to propagation model uncertainties, to test our framework. At present, we do not identify any
clear evidence for exotic contributions. However, contrasted with upcoming data on the abundance of CR nuclei,
near future measurements of the antiproton to proton flux ratio will provide a sharp probe for exotic
contributions. Such contributions could naturally arise in our model. In case that an indirect signal is observed,
measurements of the radion and KK particle masses at the LHC collider will provide a nontrivial test of the
model.

\section*{Acknowledgments}

We thank Zurab Berezhiani,  Roberto Contino, C\'edric Delaunay, Ben Gripaios, Andrey Katz, Boaz Katz,  Frank Paige, Riccardo
Rattazzi, Raman Sundrum, Tomer Volansky, Eli Waxman and Jure Zupan for useful conversations. SJL and GP also
acknowledge the Galileo Galilei Institute for Theoretical Physics in Florence, Italy, where part of this work
was completed.
KA is supported in part by NSF grant No. PHY-0652363, GP is supported by the Israel Science Foundation (grant \#1087/09), EU-FP7 Marie Curie, IRG fellowship and the Peter \&
Patricia Gruber Award.
%

\appendix

\section{Other Pati-Salam models}\label{othercust}

We present two other models with custodial symmetry for
$Z b \bar{b}$ coupling. Just like
model I (a) presented in the main text,
both these models do not
seem to fit into $SO(10)$ representations smaller than $\bf{560}$~\cite{Slansky:1981yr}.
Moreover, even if we find a fit into a suitable larger representation
of $SO(10)$, these models do not have
$SU(5)$ normalization of hypercharge
and hence might
not maintain even the SM-level of
unification of gauge couplings.

\subsection{Model I (b): : $T_{ 3 R } ^{ \nu^{ \prime } } \neq 0$
and custodial for leptons}

In Tab.~\ref{10}, we first present a model with smaller $SU(4)_c$ representations than the
benchmark model (where we had ${\bf 35}$ of $SU(4)_c$).
This model has
\begin{eqnarray}
Y & = & T_{ 3 R } + \sqrt{ \frac{8}{3} } X
\end{eqnarray}
and hence $\sin^2 \theta^{ \prime } = 3 / 11$.
Also, this model has a larger value $T_{ 3 R }^{ \nu^{ \prime } } =2$, but a smaller value of $\sin \theta^{ \prime }$
than the model with ${\bf 35}$
of $SU(4)$. Such a modification tends to enhance the DM annihilation
cross-section via $Z^{ \prime }$ exchange
into $Zh/WW$ and similarly direct detection
via $Z$ exchange, whereas it reduces the
annihilations via $Z^{ \prime }$ exchange
into top quarks (as per table~\ref{couplingscust1}).
\begin{table}[hbt]
\begin{center}
\begin{tabular}{|c|c|c|c|}
\hline
& $SU(4)_c \sim SU(3)_C \times U(1)_X$ & $SU(2)_L$ & $SU(2)_R$ \\
\hline
$t_R$, $\nu^{ \prime }$ & \bf{10} $\sim$ \bf{3}$_{ \frac{2}{3} }$,
\bf{1}$_{2}$... & \bf{1} & \bf{5}
\\
\hline
$(t,b)_L$ & \bf{10} $\sim$ \bf{3}$_{ \frac{2}{3} }$,... & \bf{2} & \bf{4}
\\
\hline
$\tau_R$ & \bf{ $\overline{4}$ } $\sim$
\bf{1}$_{ -1 }$,... & \bf{1} & \bf{1} \\
\hline
$(\nu, \tau )_L$
& \bf{ $\overline{4}$ }
$\sim$
\bf{1}$_{ -1 }$,... & \bf{2} & \bf{2} \\
\hline
$b_R$
& \bf{10}
$\sim$
\bf{3}$_{ \frac{2}{3} }$,... & \bf{1} & \bf{5} \\
\hline
\hline
$H$ & \bf{1} & \bf{2} & \bf{2} \\
\hline
\end{tabular}
\end{center}
\caption{Another model with custodial
representations for $b_L$ and RH leptons and with
non-vanishing $\nu^{ \prime } \bar{ \nu^{ \prime } } Z^{ \prime }$
coupling:  the subscripts denote
the $ \sqrt{8/3} \; X$ charge.}
\label{10}
\end{table}

\subsection{Custodial only for $b_L$, but not leptons}

Next, we present models with custodial representations only
for $b_L$ and not for RH leptons with the following two motivations in mind.
First of all, it is still interesting to have a scenario where
RH leptons are not near TeV brane and thus have
small couplings to $Z^{ \prime }$ so that we
do not need custodial representations for them. In this case,
DM annihilates mostly into hadronic SM state, i.e.,
there is no
$e^+$ signal. Moreover, we can  achieve
consistency with current $\bar{p}$ data, while simultaneously
obtaining a $\bar{p}$ signal, by simply resorting to a smaller value of SE than
what is used in order to obtain $e^+$ signal (recall that with the larger SE,
the large DM annihilation into leptons was doing a ``double-duty'' of
giving $e^+$ signal and maintaining consistency with present $\bar{p}$ data).

Moreover, there are regions of parameter space where we cannot obtain signals
from DM annihilation in cosmic rays, whether
$e^+$ (even with enhanced couplings
of $Z^{ \prime }$ to (RH) leptons) or $\bar{p}$. For example,
%
%
we can have heavy ($\sim$ TeV) DM as well as a heavy radion
so that we do not have sufficient SE.
Again, there is no motivation for custodial representations for
RH leptons in this case.
However, we still require custodial representations for $Z b \bar{b}$
so that it is still interesting to build such a
unified model.

Along these lines, the model with smallest possible representations is given
in Tab.~\ref{4},
with
\begin{eqnarray}
Y & = & T_{ 3 R } - \sqrt{ \frac{32}{3} } X
\end{eqnarray}
and hence $\sin^2 \theta^{ \prime } = 3/ 35$.
\begin{table}[hbt]
\begin{center}
\begin{tabular}{|c|c|c|c|}
\hline
& $SU(4)_c \sim SU(3)_C \times U(1)_X$ & $SU(2)_L$ & $SU(2)_R$ \\
\hline
$t_R$, $\nu^{ \prime }$ & \bf{4} $\sim$ \bf{3}$_{ \frac{-1}{3} }$,
\bf{1}$_{1}$... & \bf{1} & \bf{5}
\\
\hline
$(t,b)_L$ & \bf{4} $\sim$ \bf{3}$_{ \frac{-1}{3} }$,... & \bf{2} & \bf{4}
\\
\hline
$\tau_R$ & \bf{1} $ \sim$ \bf{1}$_{ 0 }$ & \bf{1} & \bf{3} \\
\hline
$(\nu, \tau )_L$
&
 \bf{1} $ \sim$  \bf{1}$_{ 0 }$ & \bf{2} & \bf{2} \\
\hline
$b_R$
& \bf{4}
$\sim$
\bf{3}$_{ \frac{-1}{3} }$,... & \bf{1} & \bf{5} \\
\hline
\hline
$H$ & \bf{1} & \bf{2} & \bf{2} \\
\hline
\end{tabular}
\end{center}

\caption{Simplest model with custodial
representation for $b_L$, but not for RH charged leptons:  the subscripts denote
the $ \sqrt{8/3} \; X$ charge.}
\label{4}
\end{table}

Also, this model has $T_{ 3 R } = 2$ for $\nu^{ \prime }$ and hence
is (roughly) similar to the above model with ${\bf 10}$ of $SU(4)$
as far as relic density and direct detection is concerned.
%


\section{The volume factor for antiproton propagation: a diffusion model example}\label{app:pbar}

The disc+halo diffusion model for CR propagation is widely used in the literature (see
e.g.~\cite{Ginzburg:1990sk}). In principle, the model allows to compute the CR densities arising from standard
astrophysical processes on the same footing with proposed exotic contributions, such as DM annihilation. In
practice, the model parameters are tuned on compositional CR nuclei data, which only partially constrains them.
Here we make use of this model for two purposes: (i) to clarify some issues regarding the currently fashionable
``precision treatment" of exotic CR sources within a propagation- model dependent framework, and (ii) to
illustrate the volume enhancement factor for antiprotons from a DM annihilation source, described in
Section~\ref{sssec:pbar}. Concerning the latter cause, we do not attribute particular significance to the
precise numerical results, but rather consider them as order of magnitude estimates for the expected effect.

We consider a cylindrical halo model with an infinitely thin disc, taking the diffusion coefficient as
spatially constant in the propagation volume with power law energy dependence. The model parameters relevant in
the high energy regime are $L$\,, the scale height of the cylinder, $R$\,, the radial extent, $D_0$\,, the
normalization and $\delta$\,, the power law index of the diffusion coefficient, given by
$D(\epsilon)=D_0\epsilon^\delta$\,. The parameters $L,R,D_0,\delta$ are constrained by B/C data, in such a way
as to provide the measured value of the CR grammage. For relativistic energies above a few GeV/nuc, this
constraint can be summarized as follows,
\beq\label{eq:Xdiff} X_{\rm esc}(\epsilon)\approx\frac{X_{\rm disc}Lc}{2D(\epsilon)}g(L,R)\,.\eeq
Above, $X_{\rm esc}$ is the CR grammage, $X_{\rm disc}\approx 200\,{\rm pc} \times 1.3\,m_p \times 1\,{\rm
cm}^{-3}\approx1.3\cdot10^{-3}\,{\rm gcm}^{-2}$ is the column density of the gaseous disc, where spallation
interactions occur, and c is the speed of light. The dimensionless correction factor $g(L,R)$ is given by
\beq\label{eq:g}g(L,R)=\frac{2R}{L}\sum_{k=1}^\infty J_0\left(\nu_{\rm
k}\frac{\rho_{sol}}{R}\right)\frac{\tanh\left(\nu_{\rm k}\frac{L}{R}\right)}{\nu_{\rm k}^2J_1(\nu_{\rm
k})}\,,\eeq
where $\nu_{\rm k}$ are the zeros of the Bessel function of the first kind $J_0$\,. The correction factor obeys
$g=1$ for $L\ll R$\,, and becomes smaller than one if the distance of the solar system from the radial edge is
taken comparable to the scale height of the cylinder. For the CR grammage we adopt the
parametrization~\cite{Webber:2003cj} (see also~\cite{Engelmann:1990zz,Jones:2000qd} for earlier estimates)
\beq\label{eq:X}X_{\rm esc}=27.5\epsilon^{-0.5}\,{\rm gcm}^{-2}\,.\eeq
We present the CR grammage in Eqs.~(\ref{eq:Xdiff}) and (\ref{eq:X}) as a function of energy
$\epsilon=E/\text{GeV}$\,. In fact, the grammage depends rather on magnetic rigidity, $\mathcal{R}=pc/eZ$\,.
The notation is consistent as long as we fix our attention to relativistic antiprotons.

From Eqs.~(\ref{eq:Xdiff}) and (\ref{eq:X}) we can deduce the following relation,
\beq\label{eq:DL}D_0\approx2.9\cdot10^{-2}\left(\frac{L}{4 \
\text{kpc}}\right)\tilde\epsilon^{0.5-\delta}g(L,R) \ \text{kpc}^2/\text{Myr}\,.\eeq
Equation~(\ref{eq:DL}) may now be used in order to define sets of parameters $L,R,D_0,\delta$\,, which will
agree with high energy B/C data as long as $\delta\sim0.5$\,. To this end, any high energy value of
$\tilde\epsilon\gsim10$\,GeV should do. We take $\tilde\epsilon=75$\,GeV/nuc, corresponding to the highest
energy B/C measurement by the HEAO3 mission~\cite{Stephens:1998,Jones:2000qd}. The fact that propagation (and,
in particular, diffusion) models must comply with the CR grammage is demonstrated, for example, by noting that
Eq.~(\ref{eq:DL}) holds very well for the popular MIN, MED and MAX propagation models, defined
in~\cite{Donato:2003xg} after the work of~\cite{Maurin:2001sj}.

Besides the CR grammage, additional information exists on the escape time scale, found from measurements of
radioactive CR isotopes. These data are far less accurate than the grammage measurements, and are given only
for a limited range of energies, mostly at the $\sim100$\,MeV/nuc scale~\cite{Webber:1998,Strong:2007nh}.

Different sets of values of $L,R,D_0,\delta$\,, obeying Eq.~(\ref{eq:DL}), are considered in the literature.
However we will see that, under realistic assumptions, the diffusion coefficient does not enter into the ratio
between the antiproton flux arising from DM and from the astrophysical background. In fact, to a good
approximation, the only parameter which controls this ratio is the scale height of the propagation volume. We
note at this point that, as the scale height $L$ is not independently constrained, the DM signal to
astrophysical background ratio in the disc+halo model is not constrained by the B/C data. We now proceed to
compute the flux of antiprotons resulting from DM annihilations in this propagation model example.

Neglecting losses and low energy processes and assuming steady state, the diffusion equation is
\beq\label{eq:pbdiff} -D(\epsilon)\nabla^2n=Q_{\rm DM}\,,\eeq
where $n$ is the antiproton density. The neglect of losses kept this equation easy to analyze, at the price of
moderate imprecision at energies below a few tens of GeV. We will return to this point later. Due to the
homogeneity of the diffusion coefficient, the energy dependence of the antiproton density follows that of the
source, with a trivial softening resulting from the diffusion: $n(\epsilon,\vec r)= \epsilon^{-\delta}f(\vec
r)Q(\epsilon,\vec r)$\,. We are left to deal with the spatial dependence, consisting in the function $f(\vec
r)$ for which we need to derive the value in the vicinity of the solar system.

Decomposing both $n$ and $Q$ in Bessel-Fourier series reduces the problem into an infinite set of leaky box
model-like~\cite{Ginzburg:1990sk} equations for the coefficients. We chose a decomposition in basis functions
which automatically satisfy the boundary conditions of vanishing CR density on the surface of the cylinder. For
the DM source, the decomposition reads
\beqa\begin{split}\label{eq:bessfour}Q_{\rm mk}(\epsilon)= &\frac{4}{J_1^2(\nu_{\rm
k})}\int_0^1d\zeta\cos\left[\pi\zeta\left({\rm m}+\frac{1}{2}\right)\right]
\int_0^1d\eta\eta J_0(\nu_{\rm k}\eta)Q_{\rm DM}(\epsilon,z=\zeta L,\rho=\eta R)\,,\\
Q_{\rm DM}(\epsilon,\vec r)=&\sum_{{\rm m}=0}^\infty\sum_{{\rm k}=1}^\infty Q_{\rm
mk}(\epsilon)J_0\left(\nu_{\rm k}\frac{\rho}{R}\right)\cos\left[\frac{\pi z}{L}\left({\rm
m}+\frac{1}{2}\right)\right]\,.\end{split}\eeqa
A similar decomposition holds for the antiproton density with the replacement $Q_{\rm mk}\leftrightarrow n_{\rm
mk}$\,. Using (\ref{eq:pbdiff}) we then have, for the coefficients of the antiproton density,
\beq n_{\rm mk}(\epsilon)=\frac{Q_{\rm mk}(\epsilon)L^2}{D(\epsilon)}\left[\pi^2\left({\rm
m}+\frac{1}{2}\right)^2+\nu_{\rm k}^2\frac{L^2}{R^2}\right]^{-1}\,.\eeq
Note that the DM source is separable,
\beq\label{eq:sepQpb}Q_{\rm DM}(\vec r)=Q_{{\rm DM},\bar p}(\epsilon,\vec r_{\rm sol})n_o^2(\vec r)\,, \ \ \
Q_{\rm mk}(\epsilon)=Q_{{\rm DM},\bar p}(\epsilon,\vec r_{\rm sol})q_{\rm mk}\,,\eeq
with $q_{\rm mk}$ the Bessel Fourier coefficients of $n_o^2(\vec r)$. (Recall that $n_o(\vec r)$ is defined as
the DM number density normalized to its value in the vicinity of the solar system, such that $Q_{{\rm DM},\bar
p}(\epsilon,\vec r_{\rm sol})$ is just the local injection rate due to DM.) The antiproton density in the solar
neighborhood, $z=0\,,\rho=r_{\rm sol}$\,, is thus
\beq\label{eq:npbdiff}n(\epsilon,\vec r_{\rm sol})=\frac{aL^2}{D(\epsilon)}Q_{\rm DM}(\epsilon,\vec r_{\rm
sol})\,,\eeq
with
\beq\label{eq:a}a=\sum_{{\rm m}=0}^\infty\sum_{{\rm k}=1}^\infty \frac{q_{\rm mk}J_0\left(\nu_{\rm
k}\frac{\rho_{sol}}{R}\right)}{\pi^2\left({\rm m}+\frac{1}{2}\right)^2+\nu_{\rm k}^2\frac{L^2}{R^2}}\,.\eeq

Equation~(\ref{eq:npbdiff}) allows us to obtain the specific value of the volume factor by which the DM
annihilation source is enhanced in comparison with the production by spallation. Again neglecting losses, the
antiproton density near the solar system, resulting from spallation, is~\cite{Katz:2009yd}
\beq\label{eq:npbspal} n_{\bar p,{\rm spal}}=\frac{X_{\rm esc}}{\rho_{\rm ISM}c}Q_{{\rm spal},\bar p}\,,\eeq
where $\rho_{\rm ISM}\approx1.3\,m_p\,{\rm cm}^{-3}$ is the matter density on the disc. Using
Eq.~(\ref{eq:Xdiff}) and noting that $X_{\rm disc}\approx2h\rho_{\rm ISM}$\,, where $h\sim100$\,pc is the
half-width of the disc, we obtain the ratio of the local antiproton density due to DM annihilation and due to
spallation throughout the Galaxy, expressed in terms of the local injection rates:
\beq\label{eq:pbLIBdiff}\frac{n_{\bar p,\rm DM}}{n_{\bar p,{\rm spal}}}=f_V\frac{Q_{\rm DM,\bar
p}(\epsilon,\vec r_{\rm sol})}{Q_{{\rm spal},\bar p}(\epsilon,\vec r_{\rm sol})}\,, \ \ \ \text{with} \ \ \
f_V=\frac{aL}{gh}\,.\eeq

\begin{figure}[hbp]\begin{center}
$
\begin{array}{lcr}
\includegraphics[width=90mm,height=90mm]{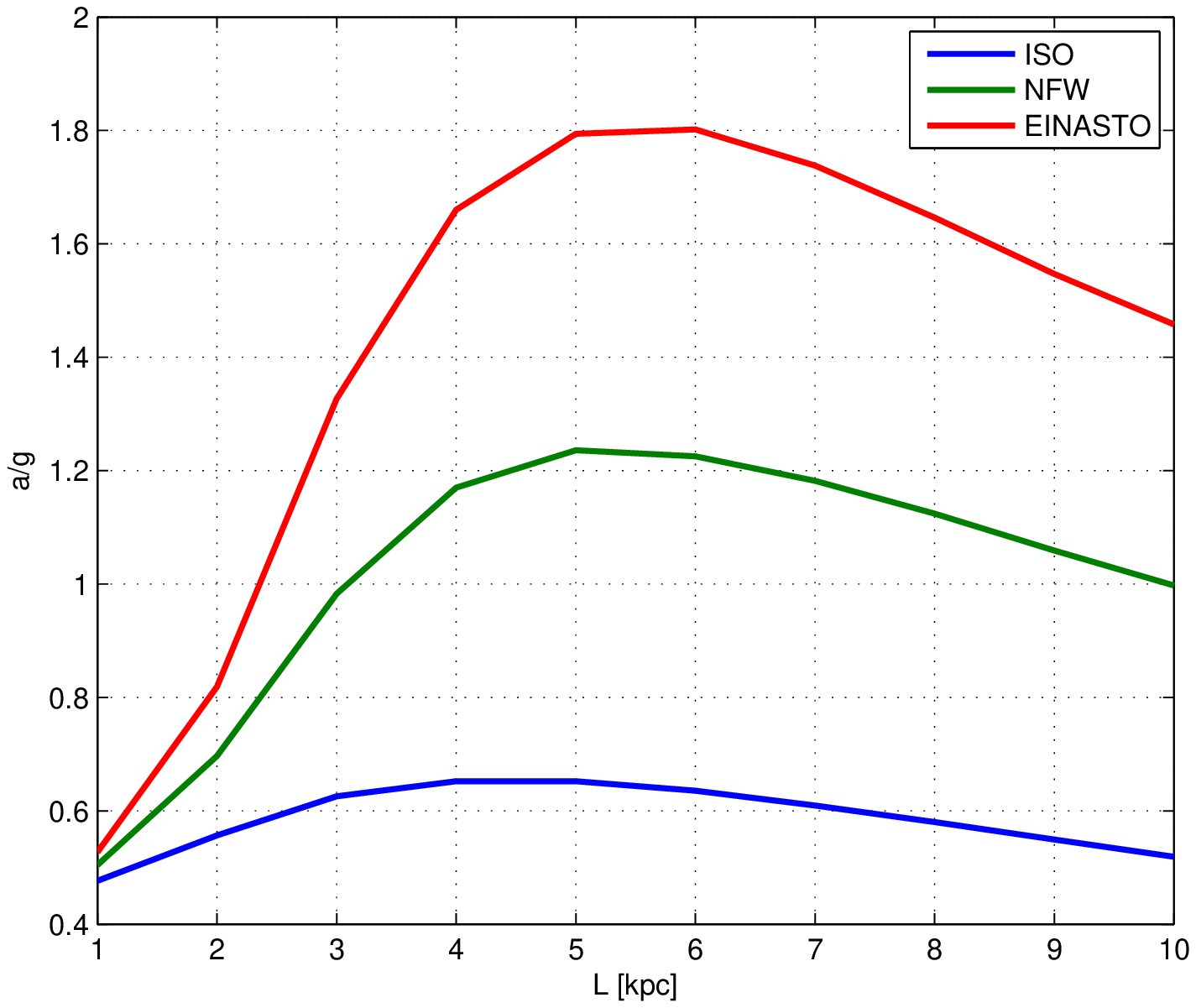}  &
\includegraphics[width=90mm,height=90mm]{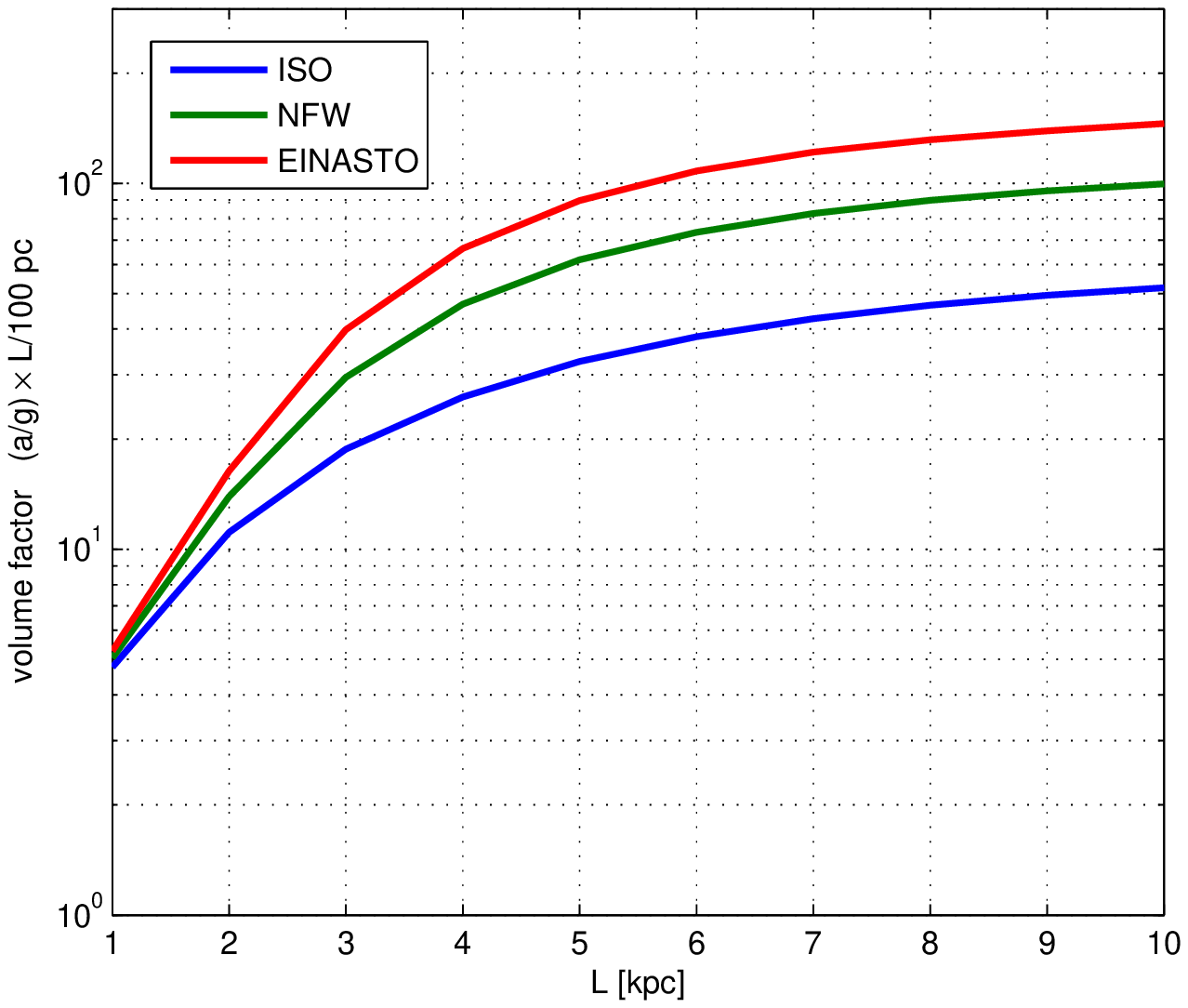}
\end{array}$
\caption{Left: The ratio between the geometrical correction factors $a$ and $g$\,, defined in the text. Right:
The volume enhancement factor for DM annihilation source over spallation source. In both panels we keep fixed
$R=20$\,kpc. On the right, we take $h=100$\,pc for the half-width of the gaseous disc.} \label{fig:aoverg}
\end{center}
\end{figure}

On the left panel of Figure~\ref{fig:aoverg} we plot the ratio of the two dimensionless correction factors
$a/g$ as a function of the CR halo half width $L$\,. We consider three DM halo profiles: the cored isothermal
(ISO) and the cusped NFW, defined in Sec.~\ref{sssec:prod}, and the Einasto profile~\cite{Merritt:2005xc}. The
ratio $a/g$ is of order unity, larger for cuspy profiles compared with the cored one. To achieve faster
convergence, we have regulated the inner cusp in the NFW and EINASTO distributions by assuming flat DM density
for $r<200$\,pc. Such inner radius is not constrained by N-body simulations. We have verified that our results
do not vary significantly as a result of increasing the regulation radius. On the right panel, we plot the
resulting volume enhancement factor $aL/gh$ for $h=100$\,pc. We find that, for reasonable values of $L$\,, the
volume factor ranges between $f_V\sim10-100$\,, depending on the assumed DM halo profile.
%

We now comment on the neglect of losses in the discussion above. For spallation antiprotons, the error due to
neglecting losses diminishes with increasing energy, as a result of the relatively rapid decrease in the
grammage. For example, the error contained in Eq.~(\ref{eq:npbspal}) due the neglect of losses is
$\approx25\%,10\% \ \text{and} \ 5\%$ at antiproton energies of $10,30 \ \text{and} \ 100$\,GeV, respectively.
Regarding the antiprotons from DM annihilation, the conclusion may be model dependent. However, in the
diffusion model considered above (as well as{\it e.g.} in the leaky box model), the escape time shares the
energy dependence of the grammage, and the conclusion is similar to the background case. In addition to losses
by collisions with ambient matter, other low energy processes are expected to influence the calculation below
the few tens of GeV. These phenomena include solar modulation, ionization losses, and even possible
reacceleration or convective motion~\cite{Strong:2007nh}. As we are dealing with a simplified propagation model
which involves, for example, ad-hoc boundary conditions for the CR halo and diffusion coefficient, and an
uncertain DM halo distribution, we find it useful to keep our expressions tractable and accurate at the high
energy $\gsim50$\,GeV regime, at the cost of minor accuracy loss below a few tens of GeV.


\end{document}